\documentclass[tbtags]{LMCS}
\def\dOi{10(1:7)2014}
\lmcsheading%
{\dOi}
{1--33}
{}
{}
{Oct.~30, 2012}
{Feb.~12, 2014}
{}

\ACMCCS{[{\bf Theory of computation}]: Formal languages and automata
  theory---Formalisms---Rewrite systems; Formal languages and automata
  theory---Tree languages}

\usepackage{ifpdf}
\ifpdf
\usepackage{pdfsync}
\fi
\usepackage[tbtags]{amsmath}
\usepackage{amssymb}
\usepackage{amsfonts}
\usepackage{mathrsfs}
\usepackage{gastex}
\usepackage{graphics}
\usepackage{xcolor}
\usepackage{latexsym}
\usepackage{ifthen}
\usepackage{url}
\usepackage{bbold}
\usepackage{pstricks,pst-node,pst-tree}
\usepackage{hyperref}

\newcommand{\Ar}{\mathsf{ranks}}

\newcommand{\bit}{\text{bit}}
\newcommand{\diff}{\mathsf{diff}}
\newcommand{\inter}{\mathsf{int}}

\newcommand{\chain}{\mathsf{chain}}
\newcommand{\lex}{\mathsf{lex}}
\newcommand{\marked}{\mathsf{marked}}
\newcommand{\marky}{\mathsf{mark}}
\newcommand{\poly}{\mathsf{poly}}
\newcommand{\NEXP}{\mathsf{NEXP}}
\newcommand{\rest}{\mathord\restriction}
\newcommand{\String}{\mathfrak{S}}
\newcommand{\size}{\mathsf{size}}
\newcommand{\sol}{\mathsf{sol}}
\newcommand{\Sol}{\mathsf{Sol}}
\newcommand{\up}{\mathsf{up}}

\newcommand{\Trans}{\mathfrak{G}}

\newcommand{\trans}[1]{\stackrel{#1}{\longrightarrow}}

\newcommand{\Ho}{\mathbb{H}}
\newcommand{\Ve}{\mathbb{V}}

\newcommand{\N}{\mathbb{N}}
\newcommand{\Z}{\mathbb{Z}}
\newcommand{\T}{\mathsf{Trees}}

\newcommand{\R}{\mathcal{R}}

\newcommand{\hook}[1]{\stackrel{#1}{\longmapsto}}

\newcommand{\problemx}[3]{
\par\noindent\underline{\sc#1}\par\nobreak\vskip.2\baselineskip
\begingroup\clubpenalty10000\widowpenalty10000
\setbox0\hbox{\bf INPUT:\ \ }\setbox1\hbox{\bf QUESTION:\ \ }
\dimen0=\wd0\ifnum\wd1>\dimen0\dimen0=\wd1\fi
\vskip-\parskip\noindent
\hbox to\dimen0{\box0\hfil}\hangindent\dimen0\hangafter1\ignorespaces#2\par
\vskip-\parskip\noindent
\hbox to\dimen0{\box1\hfil}\hangindent\dimen0\hangafter1\ignorespaces#3\par
\endgroup}

\newcommand{\problemy}[3]{
\par\noindent\underline{\sc#1}\par\nobreak\vskip.2\baselineskip
\begingroup\clubpenalty10000\widowpenalty10000
\setbox0\hbox{\bf INPUT: }\setbox1\hbox{\bf OUTPUT: }
\dimen0=\wd0\ifnum\wd1>\dimen0\dimen0=\wd1\fi
\vskip-\parskip\noindent
\hbox to\dimen0{\box0\hfil}\hangindent\dimen0\hangafter1\ignorespaces#2\par
\vskip-\parskip\noindent
\hbox to\dimen0{\box1\hfil}\hangindent\dimen0\hangafter1\ignorespaces#3\par
\endgroup}

\newcommand{\ATIME}{\mathsf{ATIME}}

\begin{document}

\title[The Complexity of the First-Order Theory of Ground Tree Rewrite Graphs]
      {The Complexity of the First-Order Theory of Ground Tree Rewrite Graphs}

\author[S.~G\"oller]{Stefan G\"oller\rsuper a}
\address{{\lsuper a}Universit\"at Bremen, Germany}
\email{goeller@informatik.uni-bremen.de}

\author[M.~Lohrey]{Markus Lohrey\rsuper b}
\address{{\lsuper b}Universit\"at Siegen, Germany}
\email{lohrey@eti.uni-siegen.de}


\begin{abstract}
The uniform first-order theory of ground tree rewrite graphs is the set of all pairs
consisting of a ground tree rewrite system and a first-order sentence that holds in 
the graph defined by the ground tree rewrite system.
We prove that the complexity of the uniform first-order theory of ground tree rewrite graphs
is in $\ATIME(2^{2^{\poly(n)}},O(n))$. 
Providing a matching lower bound, we show that there is some 
fixed ground tree rewrite graph whose first-order theory is hard
for $\ATIME(2^{2^{\poly(n)}},\poly(n))$ with respect to logspace reductions.
Finally, we prove that there exists a fixed ground tree rewrite graph together
with a single unary predicate in form of a regular tree language such that 
the resulting structure has a non-elementary first-order theory. 
\end{abstract}

\maketitle

\section{Introduction}

A ground tree rewrite system is a term rewrite system where rules 
do not contain variables (neither on the left-hand side nor on the right-hand side).
So, rules replace subtrees by subtrees.
Ground  tree rewrite systems were first studied in the term rewriting community
\cite{Bra69,DHLT90,DT90}, where they are also known as ground term rewrite systems.

Recently, ground tree rewrite systems were also studied in the context of verification
of infinite state systems \cite{loding-thesis}. The main motivation for this is that 
ground tree rewrite systems can be seen as a generalization of 
pushdown systems. These are a natural abstraction of sequential
recursive programs. Rules of a ground tree rewrite system can be applied
concurrently at different positions of a tree. This allows to model recursive progams
with  the additional ability to spawn new subthreads that are hierarchically structured,
which in turn may terminate and return some values to their parents. 

One of the most important and oldest decidability results for ground tree rewrite 
systems was shown more than 20 years ago by Dauchet and Tison \cite{DT90}:
The transition graph of a ground tree rewrite system (called a ground tree rewrite graph in the following)
has a decidable first-order theory. Actually, Dauchet and 
Tison even showed that the first-order theory of a ground tree rewrite graph extended by
the transitive closure of the edge relation is decidable (one also says that first-order logic
with reachability is decidable for ground tree rewrite graphs).
The proof of Dauchet and Tison uses a tree automata construction, which yields
a non-elementary algorithm. This leads to the question of complexity.  
While the first-order theory of a ground tree rewrite graph  extended by
the transitive closure of the edge relation may have non-elementary complexity
(this holds already for the infinite binary
tree, which is a pushdown graph \cite{Sto74}), the precise complexity of 
the first-order theories of ground tree rewrite graphs remained open.
As the  main contribution of this paper we solve
this problem. We prove the following:
\begin{itemize}
\item The first-order theory of every ground tree rewrite
graph belongs to the complexity class
$\ATIME(2^{2^{\poly(n)}},O(n))$ (doubly exponential alternating
time, where the number of alternations is bounded linearly),
where $n$ is the length of the input formula.
\item There exists a fixed ground tree rewrite graph with an
$\ATIME(2^{2^{\poly(n)}},\poly(n))$-complete
first-order theory. 
\end{itemize}
The upper bound of $\ATIME(2^{2^{\poly(n)}},O(n))$
even holds uniformly, which means that the ground tree rewrite system may be part of the
input, i.e., $n$ is the sum of the length of the input formula and the length
of the description of the ground tree rewrite system.
Let us remark that the complexity class $\ATIME(2^{2^{\poly(n)}},\poly(n))$ appears also in 
other contexts. For instance, Presburger Arithmetic (the first-order theory
of $(\N,+)$) is known to be complete for
$\ATIME(2^{2^{\poly(n)}},\poly(n))$ \cite{Ber80}, see \cite{CH90} for
similar results.

The upper bound of $\ATIME(2^{2^{\poly(n)}},O(n))$ is shown by the
method of Ferrante and Rackoff \cite{FeRa79}. Basically, the idea is to show the 
existence of a winning strategy of the duplicator in an 
Ehrenfeucht-Fra\"iss\'{e} game, where the duplicator chooses ``small''
(w.r.t. to a predefined norm) elements. This method is one of the main
tools for proving upper bounds for FO-theories. 
We divide the upper bound proof into two steps.
In a first step, we will reduce the FO-theory
for a ground tree rewrite graph to the FO-theory for a very
simple word rewrite graph, where all word rewrite rules replace one symbol by another symbol.
The alphabet will consist of all trees, whose size is bounded
by a singly exponential function in the input size (hence,
the alphabet size is doubly exponential in the input size; this
is the reason for the doubly exponential time bound).
Basically, we obtain a word over this alphabet from a tree $t$
by cutting off some upward-closed set $C$ in the tree and taking
the resulting sequence of trees. Intuitively, the set $C$ consists
of all nodes $u$ of $t$ such that the subtree rooted in $u$ is
``large''. Here, ``large'' has to be replaced by a concrete value $m
\in \N$ such that a sequence of $n$ rewrite steps applied to a tree $t$ 
cannot touch a node from the upward-closed set $C$. Clearly, $m$ depends on $n$.
In our context, $n$ will be exponential in the input size and so will $m$.
In a second step, we provide an upper bound for the FO-theory
of a word rewrite graph of the above form.

Perhaps it is worth mentioning that for proving our upper bound result one cannot make use of
Gaifman's locality theorem \cite{Gai82} since the resulting formulas in Gaifman normal form
can become non-elementary in the size of the original first-order formula \cite{DGKS07}.
An elementary upper bound on the size of Gaifman normal formulas was shown for
structures of bounded degree in \cite{DGKS07}.
However, ground tree rewrite graphs have unbounded degree.
This also the reason why Hanf's theorem \cite{Han65} does not seem to be of any use for our problem.
 
For the lower bound, we prove in a first step hardness
for $2\NEXP$ (doubly exponential nondeterministic time).
This is achieved by an encoding of a $(2^{2^n} \times 2^{2^n})$ tiling 
problem. In this tiling problem, we are given a word $w$ of length
$n$ over some fixed set of tiles, and it is asked, whether this word can 
be completed to a tiling of an array of size $(2^{2^n} \times 2^{2^n})$,
where the word $w$ is an initial part of the first row.
There exists a fixed set of tiles, for which this problem is  
$2\NEXP$-complete. From this fixed set of tiles, we construct
a fixed ground tree rewrite graph such that the following holds: From a given word 
$w$ of length $n$ over the tiles, one can construct (in logspace)
a first-order formula that evaluates to true in our fixed ground
tree rewrite graph if and only if the word $w$ is a positive
instance of the $(2^{2^n} \times 2^{2^n})$ tiling problem.
Our construction is inspired by  \cite{GoLi11},
where it is shown that the model checking problem for a fragment
of the logic EF (consisting of those EF-formulas, where on every path 
of the syntax tree at most one EF-operator occurs) over ground tree rewrite graphs is complete for 
the class $\mathsf{P}^{\NEXP}$.
In a second step, we show that our $2\NEXP$ lower bound can easily be
lifted to $\ATIME(2^{2^{\poly(n)}},\poly(n))$. For this, we have to 
consider an alternating version of the $(2^{2^n} \times 2^{2^n})$ tiling 
problem.

We conclude the paper with a proof sketch for the following result:
There exists a fixed ground tree rewrite graph together
with a single unary predicate in form of a regular tree language such that 
the resulting structure has a non-elementary first-order theory. This
result is shown by a reduction from first-order satisfiability of finite
binary words,
which is non-elementary \cite{Sto74}. It should be noted that the
first-order theory of a pushdown graph extended by regular unary
predicates still has an elementary first-order theory (it is an
automatic structure of bounded degree, hence its first-order theory
belongs to 2$\mathsf{EXPSPACE}$ by a result from \cite{KusLo09CSL}).

A short version of this paper appeared in \cite{GollerL11}.

\section{Related work}

\subsection{Other decidability and complexity results for ground tree rewrite systems}

Other important algorithmic problems that are decidable for ground tree rewrite systems are:
\begin{itemize}
\item confluence \cite{DHLT90,Oyamaguchi87}, which in fact can be decided in polynomial time \cite{ComonGN01,GodoyTV04},
\item reachability \cite{Bra69,DeruyverG89},\footnote{Actually, Brainerd \cite{Bra69} showed that a set of trees is regular if and only if it is the set
of trees that can be reached from a single tree via a ground tree rewriting system, where both translations are effective. This generalizes a result of 
B\"uchi for strings.} 
recurrent reachability \cite{loding-thesis,Loding06}, and recurrent reachability with multiple regular fairness constraints \cite{to-thesis},
\item fair termination \cite{Tis89}, and 
\item model checking certain fragments of LTL \cite{TL10,to-thesis}.
\end{itemize}
The decidability of first-order logic
with reachability for ground tree rewrite graphs implies that 
model checking of the
CTL-fragment EF is decidable for ground tree rewrite graphs; the precise complexity was 
recently shown to be non-elementary \cite{GoLi11}.

\subsection{Pushdown graphs}

As remarked above, ground tree rewrite systems generalize pushdown systems.
Muller and Schupp proved that every pushdown graph (the transition graph of a pushdown system) has a decidable monadic
second-order (MSO) theory \cite{MS85}. MSO extends first-order logic by the ability to quantify
over subsets of the universe. Most temporal logics (e.g. LTL, CTL, modal $\mu$-calculus) can be translated
into MSO and are therefore decidable over pushdown graphs. Precise complexity 
results can be found in~\cite{BEM97,MS85,Wal00,Wal01}. 

L\"oding proved in \cite{Loding02} that a ground tree rewrite graph has bounded tree width
if and only if it is a pushdown graph.  

\subsection{Algorithmic limitations}

Ground tree rewrite graphs
do not share all the nice algorithmic properties of pushdown graphs.
For instance, the infinite grid is easily seen to be (embeddable into)
a ground tree rewrite graph, which implies that  ground tree rewrite graphs with
an undecidable MSO-theory exist.
In fact, most linear-time and branching-time 
temporal logics such as LTL and CTL have undecidable model checking problems 
over ground tree rewrite graphs (cf.~\cite{loding-thesis,to-thesis}).

Concerning the first-order theory, mild generalizations of ground tree rewrite systems
lead to undecidable  first-order theories. Undecidability holds for linear and non-erasing term rewrite systems \cite{Treinen98},
right ground Noetherian rewrite systems \cite{Marcinkowski97}, and linear canonical  rewrite systems \cite{Vorobyov02}.
In all these papers, undecidability is shown for fragments of first-order logic with only one quantifier alternation.

\subsection{Formalisms related to ground tree rewrite systems}

Several other extensions of pushdown systems with multithreading capabilities have been 
considered in \cite{BMT05,KG07,Mayr-thesis,QR05}. Among these extensions,
the class of process rewrite systems \cite{Mayr-thesis}, which generalize both 
Petri nets and pushdown systems by providing hierarchical structures to 
threads, seem to have tight connections with ground tree rewrite systems. 
Lugiez and Schnoebelen proved decidability of various first-order logics on
PA-processes by using tree-automata techniques \cite{LuSc05}.
Mayr's process rewrite systems hierarchy \cite{Mayr00} was recently
refined via ground tree rewrite systems \cite{GoLi11CONCUR}.

Recently, Lin extended ground-tree rewrite systems with a finite
control unit that is acyclic but with possible self-loops, so called
weakly-extended ground tree rewrite systems \cite{Lin12}. 
It is shown that reachability, recurrent reachability and (the complement of) model checking
deterministic LTL is NP-complete for this extension.

The class of ground tree rewrite graphs is contained in the class of
 tree automatic structures \cite{Blu99}, whose FO-theories are
(non-elementarily) decidable.
 In \cite{KusLo09CSL}, it is shown that
(i) for every tree automatic structure of bounded degree (which means that the 
Gaifman-graph has bounded degree) the FO-theory belongs to 
3$\mathsf{EXPTIME}$ and that there is a fixed tree automatic structure
of bounded degree with a 3$\mathsf{EXPTIME}$-complete FO-theory.
Note that in general, ground tree rewrite graphs are {\em not} of 
bounded degree.

\subsection{Applications of the method of Ferrante and Rackoff}

Recall that  the method of Ferrante and Rackoff is the main technical tool in our proof that
the first-order theory of every ground tree rewrite
graph belongs to the complexity class
$\ATIME(2^{2^{\poly(n)}},O(n))$.
Further applications of this technique in computer
science can be found in \cite{RyVo03} (for the theory of queues)
and in \cite{Kar09} (for nested pushdown trees).

\section{Preliminaries} \label{sec:prel}

By $\Z$ we denote the {\em integers} and by $\N=\{0,1,\ldots\}$ the set of
{\em non-negative integers}. 
For $i,j\in\Z$ we define the interval $[i,j]=\{i,i+1,\ldots,j\}$ and
$[j]=[0,j]$.

For an alphabet $A$ (possibly infinite), we denote with $A^+ = A^*
\setminus \{\varepsilon\}$ the set of all non-empty words over $A$.
The length of the word $w \in A^*$ is denoted by $|w|$.
For $B \subseteq A$, we denote with $|w|_B$ the number
of occurrences of symbols from $B$ in the word $w$.

Let $f: A \to B$ be a mapping. For $A' \subseteq A$, we denote with
$f\rest A' : A' \to B$ the restriction of $f$ to $A'$.
For sets $A,B,C$ (where $A$ and $B$ may have a non-empty intersection)
and two mappings $f : A \to C$ and $g : B \to C$, we say that 
$f$ and $g$ are {\em compatible} if $f \rest (A \cap B) = g \rest (A \cap B)$.
Finally, for mappings $f : A \to C$ and $g : B \to C$
with $A \cap B = \emptyset$, we define 
$f \uplus g : A \cup B \to C$ as the mapping with 
$(f \uplus g)(a) = f(a)$ for $a \in A$ and 
$(f \uplus g)(b) = g(b)$ for $b \in B$.

\subsection{Complexity theory}

We will deal with alternating complexity classes, see \cite{CKS81,Pap-book}
for more details.  An \emph{alternating Turing-machine} is a nondeterministic
Turing-machine, where the set of states is partitioned into existential
and universal states. A configuration with a universal (resp.~existential)
state is {\em accepting} if every (resp.~some) successor configuration is accepting.
An {\em alternation} in a computation of an alternating Turing-machine is 
a transition from a universal state to an existential state or vice versa.
For functions $t(n)$ and $a(n)$ with $a(n) \leq t(n)$ for all $n
\geq 0$ let $\ATIME(t(n),a(n))$ denote the
class of all problems that can be decided on an alternating
Turing-machine in time $t(n)$ with at most $a(n)$
alternations. It is known that
$\ATIME(t(n), t(n))$ is contained in $\mbox{DSPACE}(t(n))$ if
$t(n) \geq n$ \cite{CKS81}.

\subsection{Labelled graphs}
\label{sec:labelled-graph}

A (directed) {\em graph} is a pair $(V,\rightarrow)$, where $V$ is a set of {\em
nodes} and $\rightarrow\; \subseteq V\times V$ is a binary relation.
A {\em labelled graph} is a tuple $\Trans = (V, \Sigma, \{ \xrightarrow{a} \mid a \in\Sigma\})$, 
where $V$ is a set of {\em nodes}, 
$\Sigma$ is a finite set of {\em actions}, and
$\xrightarrow{a}$ is a binary relation on $V$ for all $a
\in \Sigma$.
We note that (labelled) graphs may have infinitely many nodes.
For $u,v \in V$, we define $d_\Trans(u,v)$ as the length
of a shortest undirected path between $u$ and $v$ in the graph
$(V, \bigcup_{a \in \Sigma}\xrightarrow{a})$.
For $n \in \N$ and $u \in V$ let $S_n(\Trans,u) = \{ v \in V \mid
d_\Trans(u,v) \leq n\}$ be the {\em sphere} of radius $n$ around $u$.
Moreover, for $u_1, \ldots, u_k \in V$ 
let $S_n(\Trans,u_1,\ldots, u_k) = \bigcup_{1 \leq i \leq k} S_n(\Trans,u_i)$.
We identify  $S_n(\Trans,u_1,\ldots, u_k)$ with the substructure of $\Trans$
induced by the set $S_n(\Trans,u_1,\ldots, u_k)$, where in addition
every $u_i$ ($1 \leq i \leq k$) is added as a constant.
For two labelled graphs $\Trans_1$ and
$\Trans_2$ with node set $V_1$ and $V_2$, respectively, 
and nodes $u_1, \ldots, u_k \in V_1$,
$v_1, \ldots, v_k \in V_2$, we will consider isomorphisms
$f : S_n(\Trans_1,u_1,\ldots, u_k) \to S_n(\Trans_2,v_1,\ldots, v_k)$.
Such an isomorphism has to map $u_i$ to $v_i$. We write 
$S_n(\Trans_1,u_1,\ldots, u_k) \cong S_n(\Trans_2,v_1,\ldots, v_k)$
if there is an isomorphism 
$f : S_n(\Trans_1,u_1,\ldots, u_k) \to S_n(\Trans_2,v_1,\ldots, v_k)$.

\begin{lem} \label{lemma:disjoint-spheres}
Let $\Trans_1, \Trans_2$ be labelled graphs with the same set of
actions and node sets $V_1$ and $V_2$, respectively. 
Let $\overline{u} \in V_1^k$, $\overline{v} \in V_2^k$,
$u \in V_1$, and $v \in V_2$ such that
$u \not\in S_{2n+1}(\Trans_1,\overline{u})$ and 
$v \not\in S_{2n+1}(\Trans_2,\overline{v})$.
Finally, let
$f : S_n(\Trans_1, \overline{u}) \to S_n(\Trans_2, \overline{v})$
and $f' : S_n(\Trans_1, u) \to S_n(\Trans_2, v)$ be isomorphisms.
Then $f \uplus f' : S_n(\Trans_1, u, \overline{u}) \to S_n(\Trans_2, v,
\overline{v})$ is an isomorphism as well.
\end{lem}

\begin{proof}
The lemma is obvious, once one realizes that 
the condition $u \not\in S_{2n+1}(\Trans_1,\overline{u})$
implies that the spheres $S_n(\Trans_1,\overline{u})$
and $S_n(\Trans_1, u)$ are disjoint and that there is no edge
between the two spheres (and similarly for the spheres
$S_n(\Trans_2,\overline{v})$ and $S_n(\Trans_2, v)$).
\end{proof}

Later, we have to lift a relation $\to$ on a set $A$ to a larger set.
We will denote this new relation again by $\to$. 
Two constructions will be needed.
Assume that $\to$ is a binary relation on a set $A$
and let $A \subseteq B$.
We lift $\to$ to 
the set $B^+$ of non-empty words over $B$ as follows: 
For all $u, v \in B^+$, we have $u \to v$ if and only if there 
are $x,y \in B^*$ and $a,b \in A$ such that $a \to b$ and 
$u  = xay$, $v = xby$. Note that this implies $|u|=|v|$.
The second construction lifts $\to \; \subseteq A \times A$ 
from $A$ to $\N \times A$ as follows:
For $a,b \in A$ and $m,n \in \N$ let $(m,a) \to (n,b)$ if and only
if $m=n$ and $a \to b$. 
Note that $(\N \times A, \to)$ consists of $\aleph_0$ many
disjoint copies of $(A, \to)$. Moreover, 
$((A \cup \{\$\})^+ \setminus\{\$\}^+, \to)$ (where $\$ \not\in A$ is a new symbol)
is isomorphic to 
$(\N \times A^+, \to)$.
 
 \begin{exa} \label{example extension to words}
 For the relation $\to \ = \ \{ (a,b), (b,a) \}$ the corresponding relation on $\{a,b\}^+$ is shown in Figure~\ref{fig-exa-word-graph}.
 The relation $\to$ lifted to $\N \times \{a,b\}$ is simply the disjoint union of all 2-cycles
 \begin{center}                                                                               
\setlength{\unitlength}{1.5mm}                                                               
\begin{picture}(10,4)(0,-2)                                                                  
\gasset{ELdist=0.2,linewidth=.15,ExtNL=n,Nframe=n,Nadjust=wh,Nadjustdist=0.3}                
\node(a)(0,0){$(a,n)$}
\node(b)(10,0){$(b,n)$}
\drawedge[curvedepth=1](a,b){}
\drawedge[curvedepth=1](b,a){}
\end{picture}
\end{center} 
for all $n \in \N$.
 \end{exa}
\begin{figure}[t]                                                                            
\begin{center}                                                                               
\setlength{\unitlength}{1.5mm}                                                               
\begin{picture}(53,15)                                                                  
\gasset{ELdist=0.2,linewidth=.15,ExtNL=n,Nframe=n,Nadjust=wh,Nadjustdist=0.3}                
\node(a)(0,0){$a$}
\node(b)(0,10){$b$}
\drawedge[curvedepth=1](a,b){}
\drawedge[curvedepth=1](b,a){}
\node(aa)(10,0){$aa$}
\node(ab)(10,10){$ab$}
\node(bb)(20,10){$bb$}
\node(ba)(20,0){$ba$}
\drawedge[curvedepth=1](aa,ab){}
\drawedge[curvedepth=1](ab,aa){}
\drawedge[curvedepth=1](ab,bb){}
\drawedge[curvedepth=1](bb,ab){}
\drawedge[curvedepth=1](bb,ba){}
\drawedge[curvedepth=1](ba,bb){}
\drawedge[curvedepth=1](ba,aa){}
\drawedge[curvedepth=1](aa,ba){}

\node(aaa)(30,0){$aaa$}
\node(aab)(30,10){$aab$}
\node(abb)(40,10){$abb$}
\node(aba)(40,0){$aba$}
\drawedge[curvedepth=1](aaa,aab){}
\drawedge[curvedepth=1](aab,aaa){}
\drawedge[curvedepth=1](aab,abb){}
\drawedge[curvedepth=1](abb,aab){}
\drawedge[curvedepth=1](abb,aba){}
\drawedge[curvedepth=1](aba,abb){}
\drawedge[curvedepth=1](aba,aaa){}
\drawedge[curvedepth=1](aaa,aba){}
\node(baa)(35,5){$baa$}
\node(bab)(35,15){$bab$}
\node(bbb)(45,15){$bbb$}
\node(bba)(45,5){$bba$}
\drawedge[curvedepth=1](baa,bab){}
\drawedge[curvedepth=1](bab,baa){}
\drawedge[curvedepth=1](bab,bbb){}
\drawedge[curvedepth=1](bbb,bab){}
\drawedge[curvedepth=1](bbb,bba){}
\drawedge[curvedepth=1](bba,bbb){}
\drawedge[curvedepth=1](bba,baa){}
\drawedge[curvedepth=1](baa,bba){}
\drawedge[curvedepth=1](aaa,baa){}
\drawedge[curvedepth=1](baa,aaa){}
\drawedge[curvedepth=1](aab,bab){}
\drawedge[curvedepth=1](bab,aab){}
\drawedge[curvedepth=1](aba,bba){}
\drawedge[curvedepth=1](bba,aba){}
\drawedge[curvedepth=1](abb,bbb){}
\drawedge[curvedepth=1](bbb,abb){}
\put(50,7.5){$\ldots$}
\end{picture}
\end{center} 
\caption{\label{fig-exa-word-graph} A finite portion of the relation $\to$ from Example~\ref{example extension to words} extended to $\{a,b\}^+$.}
\end{figure}
 
For a labelled graph
$\Trans = (V, \Sigma, \{ \xrightarrow{a} \mid a \in\Sigma\})$,
we define the labelled graph
\begin{equation} \label{def-G^+}
\Trans^+ = (V^+,\Sigma, \{ \xrightarrow{a} \mid a \in\Sigma\}).
\end{equation}
Note that by the above definition, $\xrightarrow{a}$ is lifted
to a relation on $V^+$.

\subsection{First-order logic}
\label{sec:FO}

We will consider first-order logic (briefly FO) with equality over labelled graphs.
Thus, for a set $\Sigma$ of actions, we have for each $a \in \Sigma$
a binary relation symbol $a(x,y)$ in our signature. The meaning 
of $a(x,y)$ is of course $x \trans{a} y$. 
If $\varphi(x_1, \ldots, x_n)$ is a first-order formula with free variables
$x_1, \ldots, x_n$, $\Trans = (V, \Sigma, \{ \xrightarrow{a} \mid a \in\Sigma\})$
is a labelled graph, and $v_1, \ldots, v_n \in V$, then we write 
$\Trans \models \varphi(v_1, \ldots, v_n)$ if $\varphi$ evaluates to true
in $\Trans$, when variable $x_i$ is instantiated by $v_i$ ($1 \leq i \leq n$).
The {\em first-order theory} (briefly {\em FO-theory}) 
of a labelled transition graph $\Trans$ is the set of all first-order sentences
(i.e., first-order formulas without free variables) $\varphi$ with 
$\Trans\models\varphi$. In the final Section~\ref{sec:GTRS+unary}, we will consider
the first-order theory of a labelled graph with an additional unary predicate.
The {\em quantifier rank} of a first-order formula is the maximal number 
of nested quantifiers in $\varphi$.
We will need the following well known lemma, which goes back to work of
Fischer and Rabin \cite{Fischer74super-exponentialcomplexity}.

\begin{lem} \label{lemma:Fischer-Rabin}
Let $\Sigma$ be a set of actions. Given
a first-order formula $\theta(x,y)$ of quantifier rank $\mathsf{qr}(\theta)$
and a binary-coded integer $j$ (let $m$ be the number of $1$-bits
in the binary representation of $j$), one can compute in logspace
a first-order formula $\theta^j(x,y)$ of 
quantifier rank $O(\log(j) + \mathsf{qr}(\theta))$ and
size $O(m \cdot \log(j) + m \cdot |\theta|)$ 
such that for every labelled graph $\Trans = (V, \Sigma, \{ \xrightarrow{a} \mid a \in\Sigma\})$
and all nodes $u,v \in V$ we have: 
$\Trans \models \theta^j(u,v)$ if and only if there is a directed
path of length $j$ from $u$ to $v$ in the graph $(V, \{ (s,t) \mid
\Trans \models \theta(s,t) \})$.  
\end{lem}

\begin{proof}
Before we define
$\theta^j(x,y)$, let us inductively define for each $k\in \N$ a formula
$\psi_k(x,y)$ such that for all $u,v\in V$ we have
$\Trans \models \psi_k(u,v)$ if and only if 
there is a directed
path of length $2^k$ from $u$ to $v$ in the graph $(V, \{ (s,t) \mid
\Trans \models \theta(s,t) \})$.  
 We define
\begin{eqnarray*}
\psi_0(x,y)&=& \theta(x,y), \text{ and}\\
\psi_k(x,y)&=& \exists z\forall u,v \bigg(
\big((u=x\wedge v=z)\vee(u=z\wedge v=y)\big)\rightarrow
\psi_{k-1}(u,v) \bigg) \text{ for } k \geq 1.
\end{eqnarray*}
Note that the size of $\psi_k(x,y)$ is $ O(k + |\theta|)$ and the
quantifier rank is $3 k + \mathsf{qr}(\theta)$.

Let $U\subseteq \N$ be the set of all positions of the binary representation
of $j$ whose bit is set to $1$, i.e., $j=\sum_{i\in U}2^i$.
Let $m=|U|$ and let $h_1, \ldots, h_m$ be some enumeration of $U$.
We can now define $\theta^j(x,y)$ as 
$$\theta^j(x,y)\quad=\quad\exists x_1,\ldots,x_{m+1} \bigg( x_1=x
  \ \wedge\ x_{m+1}=y \ \wedge \
  \bigwedge_{i\in[1,m]}\psi_{h_i}(x_i,x_{i+1})\bigg).$$
From the binary representation of $j$, we can easily compute
$\theta^j(x,y)$.
Moreover, the size of $\theta^j(x,y)$ is bounded
by $O(m \cdot \log(j) + m \cdot |\theta|)$ and 
the quantifier rank is bounded by $O(\log(j) + \mathsf{qr}(\theta))$.
\end{proof}

One of most successful techniques for proving upper bounds for the 
complexity of first-order theories is the method of Ferrante and
Rackoff \cite{FeRa79}. We will apply this method in Section~\ref{sec:string}.
The following result is shown in 
\cite{FeRa79}.\footnote{The actual statement in \cite{FeRa79} is
  stronger, but for our purpose the weaker statement in Theorem~\ref{thm-FR} is sufficient.}

\begin{thm} \label{thm-FR}
Let $\Trans$ be a labelled graph, and let $V$ be the set
of nodes of $\Trans$. Assume that for every node $v \in V$ we have
a norm $|v| \in \N$ (in our application, $V$ will be a set of words and 
the norm of a word will be its length). Let $V_n = \{ v\in V \mid |v|
\leq n\}$. Moreover, for $k,\ell \geq 0$, let $\equiv_{k,\ell}$ be 
an equivalence relation on the set $V^k$ and let $H : \N^2 \to \N$ 
be a function such that the following  
properties hold for all $k,\ell \in \N$,
$\overline{u}, \overline{v} \in V^k$:
\begin{enumerate}[label=\({\alph*}]
\item If $\overline{u} \equiv_{k,0} \overline{v}$, then 
$\overline{u}$ and $\overline{v}$ satisfy the same quantifier-free
formulas in the structure $\Trans$.
\item If $\overline{u} \equiv_{k,\ell} \overline{v}$ and $\ell>0$, 
 then for all $u \in V$
 there exists $v \in V_{H(k,\ell)}$ with
 $(\overline{u},u) \equiv_{k+1,\ell-1} (\overline{v},v)$.
\end{enumerate}
Then, for every quantifier-free formula
$\psi(x_0, \ldots, x_\ell)$ and all quantifiers $Q_0, \ldots, Q_\ell
\in \{\exists, \forall\}$ we have that $\Trans \models Q_0 x_0 \cdots
Q_\ell x_\ell : \psi(x_0, \ldots, x_\ell)$ if and only if
$$
\Trans \models Q_0 x_0 \in V_{H(0,\ell)} 
Q_1 x_1 \in V_{H(1,\ell-1)} 
\cdots
Q_\ell x_\ell \in V_{H(\ell,0)} : \psi(x_0, \ldots, x_\ell).
$$
\end{thm}
We will use Theorem~\ref{thm-FR}
in Section~\ref{sec:string}, where the function $H(k,\ell)$ will be exponential in $k+\ell$.

\subsection{Trees}
\label{sec:trees}

Let $\preceq$ denote the prefix order on $\N^*$, i.e., $x\preceq y$ for
$x,y\in\N^*$ if there is some $z\in\N^*$ such that $y=xz$. 
A set $D \subseteq \N^*$ is called {\em prefix-closed}
if for all $x,y \in \N^*$, $x \preceq y \in D$ implies $x \in D$.
A {\em ranked alphabet} is a collection of finite and pairwise disjoint
alphabets $A=(A_i)_{i\in[k]}$ for some $k\geq 0$
such that $A_0 \neq \emptyset$.
For simplicity we identify $A$ with $\bigcup_{i\in[k]}A_i$.
A {\em ranked tree} (over the ranked alphabet $A$) is a 
mapping $t:D_t\rightarrow A$, where $D_t\subseteq[1,k]^*$ satisfies the following:
\begin{itemize}
\item $D_t$ is non-empty, finite, and prefix-closed, and
\item for each $x\in D_t$ with $t(x)\in A_i$ we have $x1,\ldots,xi\in D_t$
and $xj\not\in D_t$ for each $j>i$.
\end{itemize}
We say that $D_t$ is the {\em domain} of $t$ and call its elements {\em nodes}.
In case $t(x)\in A_2$ for some node $x$, then $x1$ is the {\em left child}
and $x2$ the {\em right child} of $x$. 
A {\em leaf} of $t$ is a node $x$ with $t(x)\in A_0$.
An {\em internal node} of $t$ is a node, which is not a leaf.
We also refer to $\varepsilon\in D_t$ as the {\em root} of $t$.
By $\T_A$ we denote the set of all ranked trees over the ranked alphabet $A$. 
Define $\size(t)$ as the number of nodes in a tree $t$. 
It is easy to show that the number of all trees from $\T_A$ of size 
at most $n$ is bounded by $|A|^n$.

\begin{exa}
Assume $A_0 = \{a,b\}$, $A_1=\{g\}$, and $A_2 = \{f\}$.
Figure~\ref{fig-tree} shows a tree $s \in \T_A$ with 
$\size(s)=11$. The domain $D_s$ of this tree is
$$
\{\varepsilon,1,2,11,12,21,22,111,121,1211,221\}.
$$
\end{exa}
\begin{figure}[t]                                                                            
\begin{center}                                                                               
\setlength{\unitlength}{1.1mm}                                                               
\begin{picture}(60,34)(0,26)                                                                  
\gasset{AHnb=0,ELdist=0.2,linewidth=.15,ExtNL=n,Nframe=n,Nadjust=wh,Nadjustdist=0.7}                
\node(r)(30,60){$f$}                                                                         
\node(1)(20,50){$f$}
\node(2)(40,50){$f$}
\node(11)(13,40){$g$}
\node(12)(27,40){$g$}
\node(21)(33,40){$a$}
\node(22)(47,40){$g$}
\node(111)(13,33){$a$}
\node(121)(27,33){$g$}
\node(1211)(27,26){$b$}
\node(221)(47,33){$a$}                                         
\drawedge(r,1){} 
\drawedge(r,2){} 
\drawedge(1,11){} 
\drawedge(1,12){} 
\drawedge(2,21){} 
\drawedge(2,22){} 
\drawedge(11,111){}
\drawedge(12,121){}
\drawedge(121,1211){}
\drawedge(22,221){}
\end{picture}
\end{center}
\caption{\label{fig-tree} A tree $s$}
\end{figure}
Let $t$ be a ranked tree and let $x$ be a node of $t$.
For each $x\in[1,k]^*$ we define $xD_t=\{xy\in[1,k]^*\mid y\in D_t\}$
and $x^{-1}D_t=\{y\in[1,k]^*\mid xy\in D_t\}$.
By $t^{\downarrow x}$ we denote the {\em subtree of $t$} 
with root $x$, i.e., the tree with domain
$D_{t^{\downarrow x}}=x^{-1}D_t$
defined as $t^{\downarrow x}(y)=t(xy)$.
Let $s,t\in\T_A$ and let $x$ be a node of $t$.
We define $t[x/s]$ to be the tree that is obtained by replacing
$t^{\downarrow x}$ in $t$ by $s$, more formally
$D_{t[x/s]}=(D_t\setminus xD_{t^{\downarrow x}})\cup xD_s$
with 
$$
t[x/s](y)=
\begin{cases}
t(y)&
\quad\text{ if } y\in D_t\setminus xD_{t^{\downarrow x}}\\
s(z)&
\quad\text{ if } y=xz\text{ with }z\in D_s.
\end{cases}
$$
For two ranked trees $s$ and $t$, let $\diff(s,t) = |D_s \setminus D_t|$.
Thus $\diff(s,t)$ is the number of nodes that belong to the tree $s$
but not to the tree $t$.

\begin{exa}
Consider the tree $s$ from Figure~\ref{fig-tree} and the 
tree $t$ from Figure~\ref{fig-chain}. We have 
$$
D_s \setminus D_t = \{11,12,22,111,121,1211,221\}
$$
and hence $\diff(s,t)=7$.
\end{exa}
\begin{figure}[t]                                                                            
\begin{center}                                                                               
\setlength{\unitlength}{1.1mm}                                                               
\begin{picture}(60,35)(0,25)                                                                  
\gasset{AHnb=0,ELdist=0.2,linewidth=.15,ExtNL=n,Nframe=n,Nadjust=wh,Nadjustdist=0.7}                
\node(r)(30,60){$f$}                                                                         
\node(1)(25,53){$a$}
\node(2)(37,53){$g$}
\node(21)(37,46){$f$}
\node(211)(30,39){$g$}
\node(212)(44,39){$a$}
\node(2111)(30,32){$g$}
\node(21111)(30,25){$b$}           
\drawedge(r,1){} 
\drawedge(r,2){} 
\drawedge(2,21){} 
\drawedge(21,211){}
\drawedge(21,212){}
\drawedge(211,2111){}
\drawedge(2111,21111){}
\end{picture}
\end{center}
\caption{\label{fig-chain} A chain $t$}
\end{figure}
Let $C$ be a prefix-closed subset of $D_t$. 
We define the string of subtrees $t \setminus C$ as follows:
If $C = \emptyset$, then $t \setminus C = t$. If $C \neq \emptyset$,
then $t \setminus C = t^{\downarrow v_1} \cdots t^{\downarrow v_m}$, where
$v_1, \ldots, v_m$ is a list of all nodes from $((C \cdot \N) \cap D_t) \setminus C$ in
lexicographic order. Intuitively, we remove from the tree $t$ the 
prefix-closed subset $C$ and list all remaining maximal subtrees.
For $n \in \mathbb{N}$ and a tree $t$
we define the 
prefix-closed subset $\up(t,n)\subseteq D_t$ as 
$$
\up(t,n) = \{ v \in D_t \mid \size(t^{\downarrow v}) > n \}.
$$
Note that $t \setminus \up(t,n)$ is a list of all maximal subtrees
of size at most $n$ in $t$, listed in lexicographic order.

\begin{exa}
Consider the tree $s$ from 
Figure~\ref{fig-tree}. Then 
$$C = \{\varepsilon,1,2,12\} \subseteq D_s$$
is prefix-closed. We have
$$
s \setminus C = g(a),g(b),a,g(a)
$$
(here, we denote trees by their corresponding term expressions, and we
separate the trees in the sequence $s \setminus C$ with the
symbol ``,''). Moreover, we have $C = \up(s,2)$.
\end{exa}
A tree $t \in \T_A$
is a {\em chain} if $D_t \neq \{\varepsilon\}$ and
for every internal node $u \in D_t$,
there is at most one child $ui$ of $u$ such that $ui$
is internal.  
Hence, a chain $t$ has
a unique maximal (with respect to the prefix relation) internal node
$\max(t) \in \N^*$.  Note that a chain consists of at least two nodes.

\begin{exa}
The tree $t$ in Figure~\ref{fig-chain} is a chain with $\max(t)=2111$.
\end{exa}

\begin{lem} \label{lemma:number-leaves}
Let $A$ be a ranked alphabet and let $\Ar = \{ m \in \N \mid m \geq 1, A_m \neq \emptyset \}$.
Then, for all $n \geq 1$, the following are equivalent:
\begin{enumerate}[label=\({\alph*}]
\item  There is a chain $t \in \T_A$ with exactly 
$n$ leaves.
\item  There is a tree $t \in \T_A$ with exactly
$n$ leaves.
\item  There exist numbers $d_m \in \N$ (for each $m \in \Ar$) such that
$n = 1 + \sum_{m \in \Ar} d_m \cdot (m-1)$.  
\end{enumerate}
\end{lem}

\begin{proof}
Implication $(a) \Rightarrow (b)$ is trivial.
Now, assume (b) and let $t \in \T_A$ has exactly $n$ leaves.
We show (c)  by induction on the size 
of $t$. We distinguish two cases. The case $n = 1$ is clear; set
$d_m=0$ for all $m \in \Ar$. Now, assume that $t$ has 
$n \geq 2$ leaves. Then, there must exist an internal node $u \in D_t$
such that all children of $u$ are leaves. Let $1 \leq a \leq n$ be the rank of 
the symbol $t(u)$. By replacing $u$ by a leaf (labelled with an
arbitrary constant from $A_0$), we get a strictly smaller tree with 
$n - (a-1)$ many leaves (note that $a=1$ is possible). Since $a \leq
n$  we have $n - (a-1) \geq 1$.
By induction, there exist 
$d_m \in \N$ ($m \in \Ar$) such that
$n - (a-1) = 1 + \sum_{m \in M} d_m \cdot (m-1)$.
Thus, we have
$n = 1 + (d_a+1) \cdot (a-1) + \sum_{m \in \Ar\setminus\{a\}} d_m \cdot
(m-1)$.

Finally, for the implication $(c) \Rightarrow (a)$, assume that 
$n = 1 + \sum_{m \in \Ar} d_m \cdot (m-1)$. 
Take a chain $t$ that consists of $\sum_{m \in \Ar} d_m$ 
internal nodes, $d_m$ of which are labelled with a symbol of
rank $m$. All other nodes are leaves. It is a simple observation
that $t$ has exactly $n$ leaves. 
\end{proof}
The following lemma follows directly from
Lemma~\ref{lemma:number-leaves}.

\begin{lem} \label{lemma:number-leaves(2)}
Let $A$ be a ranked alphabet and let $\Ar = \{ m \in \N \mid m \geq 1, A_m \neq \emptyset \}$.
Then, for every tree $t \in \T_A$ and every prefix-closed subset
$C$ of $D_t$ the following holds, where $n$ is the length of the string
$t \setminus C$:
There exist  numbers $d_m \in \N$ (for each $m \in \Ar$) such that
$n = 1 + \sum_{m \in \Ar} d_m \cdot (m-1)$. 
\end{lem}

\subsection{Ground tree rewrite graphs}

A {\em ground tree rewrite system (GTRS)} is tuple
$\R=(A,\Sigma,R)$, where $A$ is a ranked alphabet,
$\Sigma$ is finite set of actions,
and $\R \subseteq \T_A \times \Sigma \times \T_A$
is a finite set of rewrite rules. A rule
$(s,a,s')$ is also written as 
$s\hook{a}s'$.
The {\em ground tree rewrite graph} defined by $\R$ is 
$$\Trans(\R)=(\T_A,\Sigma,\{\trans{a}\mid a\in\Sigma\}),
$$ 
where for each $a\in\Sigma$, we have
$t\trans{a}t'$ if and only if there exist a rule 
$(s\hook{a}s') \in R$ and $x\in D_t$
such that $t^{\downarrow x}=s$ and $t'=t[x/s']$.

\begin{exa}
We define a GTRS $\R=(A,\Sigma,R)$ as follows.
Let $A_0 = \{a,b\}$, $A_1=\{g\}$, and $A_2 = \{f\}$,
$\Sigma=\{a,b\}$, and let $R$ consist of the following two rules:
$$
a \hook{a} g(a), \qquad b \hook{b} g(b) .
$$
Take a tree $t(a_1,a_2, \ldots, a_n)$, where $a_1, \ldots, a_n \in \{a,b\}$, that 
does not contain a subtree of the form $g(a)$ or $g(b)$.
Then, the (weakly) connected component of $\Trans(\R)$ that contains
$t(a_1,a_2, \ldots, a_n)$ consists of all trees of the form $t(g^{i_1}(a_1),g^{i_2}(a_2), \ldots, g^{i_n}(a_n))$
for $i_1, \ldots, i_n \geq 0$. These trees form an $n$-dimensional grid, where edges in dimension $1 \leq j \leq k$ 
are labelled with $a_j$.
Figure~\ref{fig-grid} shows the connected component of $\Trans(\R)$ that contains $f(a,b)$.
\end{exa}
\begin{figure}[t]                                                                            
\begin{center}                                                                               
\setlength{\unitlength}{1mm}                                                               
\begin{picture}(60,60)                                                                  
\gasset{ELdist=0.2,linewidth=.15,ExtNL=n,Nframe=n,Nadjust=wh,Nadjustdist=0.7}                
\node(00)(0,0){$\scriptstyle f(a,b)$}                                                                         
\node(01)(30,0){$\scriptstyle f(a,g(b))$}
\node(02)(60,0){$\scriptstyle f(a,g(g(b)))$}
\node(10)(0,30){$\scriptstyle f(g(a),b)$}
\node(11)(30,30){$\scriptstyle f(g(a),g(b))$}    
\node(12)(60,30){$\scriptstyle f(g(a),g(g(b)))$}  
\node(20)(0,60){$\scriptstyle f(g(g(a)),b)$}
\node(21)(30,60){$\scriptstyle f(g(g(a)),g(b))$}    
\node(22)(60,60){$\scriptstyle f(g(g(a)),g(g(b)))$}  

\drawedge(00,01){$b$} 
\drawedge(10,11){$b$}
\drawedge(20,21){$b$}
\drawedge(01,02){$b$} 
\drawedge(11,12){$b$}
\drawedge(21,22){$b$}
\drawedge(00,10){$a$}
\drawedge(01,11){$a$}
\drawedge(02,12){$a$}
\drawedge(10,20){$a$}
\drawedge(11,21){$a$}
\drawedge(12,22){$a$}
\end{picture}
\end{center}
\caption{\label{fig-grid} A finite part of the graph $\Trans(\R)$}
\end{figure}
The next two lemmas are obvious:

\begin{lem}\label{lemma-dist}
Let $\R = (A,\Sigma,R)$ be a GTRS and let $r$ be 
the maximal size of a tree that appears in $R$.
Let $s$ and $t$ be ranked trees such that 
$d_{\Trans(\R)}(s,t) \leq n$. Then $\size(t) \leq \size(s) + r \cdot n$.
\end{lem}

\begin{lem}\label{lemma-diff}
Let $\R = (A,\Sigma,R)$ be a GTRS and let $r$ be
the maximal size of a tree that appears in $R$.
Let $s$ and $t$ be ranked trees such that $\diff(s,t) > r \cdot n$.
Then $d_{\Trans(\R)}(s,t) > n$.
\end{lem}
Recall the definition of the graph $\Trans^+$ from
\eqref{def-G^+}.

\begin{lem}\label{lemma-GTRS-spheres}
Let $\R = (A,\Sigma,R)$ be a GTRS and let $r$ 
be the maximal size of a tree that appears in $R$.
Let $t$ be a ranked tree, $n \in \N$, and let $C \subseteq 
\up(t, r \cdot n)$ be prefix-closed. 
Then we have
$$
S_n(\Trans(\R), t) \cong S_n(\Trans(\R)^+, t \setminus C).
$$ 
\end{lem}

\begin{proof}
Let $t \setminus C = t_1 \cdots t_m$.
Hence, there is a tree $s$ with $m$ leaves such that
$t$ results from $s$ by replacing the $i^{\text{th}}$ leaf of $s$ by $t_i$ ($1 \leq i \leq m$),
let us write $t = s[t_1, t_2, \ldots, t_m]$ for this. 
Recall that the subtree rooted in a node from $C \subseteq \up(t, r \cdot n)$ has size 
strictly larger than $r \cdot n$.  Therefore, a node from $C$ cannot 
be accessed by doing at most $n$ rewrite steps. 
Hence, every tree $t' \in S_n(\Trans(\R), t)$ can be written (uniquely) as
$t' = s[t_1', t_2', \ldots, t_m']$.
Moreover, the mapping $t' \mapsto t_1' t_2' \cdots t_m'$
defines an isomorphism from $S_n(\Trans(\R), t)$ to $S_n(\Trans(\R)^+,
t \setminus C)$.
\end{proof}

\begin{rem} \label{rem-GTRS-spheres}
Note that if the word $w \in \T_A^+$ results from the string $t \setminus C$ by permuting the 
trees in the string, then we still have $S_n(\Trans(\R), t) \cong S_n(\Trans(\R)^+, w)$.
\end{rem}
The main goal of this paper is to study the complexity of  the following set that we call the
{\em uniform first-order theory of  ground tree rewrite graphs}:
\begin{align*}
\{ (\R, \varphi) \mid \  & \R=(A,\Sigma,R) \text{ is a GTRS, } \varphi \text{ is an FO-sentence over the signature of }  \Trans(\R),  \\
& \Trans(\R)\models\varphi \}.
\end{align*}


\renewcommand{\O}{\mathbb{1}}
\renewcommand{\Z}{\mathbb{O}}

\section{An $\ATIME(2^{2^{\poly(n)}}, O(n))$ upper bound}

In this section we will prove the following result:

\begin{thm} \label{thm:upper-main}
The uniform first-order theory of  ground tree rewrite graphs
belongs to the complexity class $\ATIME(2^{2^{\poly(n)}},O(n))$.
\end{thm}
It suffices to prove Theorem~\ref{thm:upper-main} for the case 
that the underlying ranked alphabet $A$ contains a symbol
of rank at least two. A ground tree rewrite graph, where 
all symbols have rank at most 1 is in fact a suffix rewrite graph on words.
Such a graph is first-order interpretable in a full
$|\Gamma|$-ary tree $\Gamma^*$ (with $\Gamma$ finite), where 
the defining first-order formulas can be easily computed from
the suffix rewrite system.
Finally, the first-order theory of a full tree $\Gamma^*$ 
(with $|\Gamma| \geq 2$) is complete
for the class $\ATIME(2^{O(n)}, O(n))$ (under 
log-lin reductions) \cite{CH90,Vogel83}. 

The proof of Theorem~\ref{thm:upper-main} will be divided
into two steps. In a first step, we will reduce the FO-theory
for a given ground tree rewrite graph to the FO-theory for a very
simple word rewrite graph of the form $\Trans^+$, where $\Trans$
is a finite labelled graph. Note that if $V$ is the set of nodes
of $\Trans$, then $V^+$ is the set of nodes of $\Trans^+$.
Moreover, every edge in $\Trans^+$ replaces a single symbol in a word
by another symbol. 
In our reduction, the size of the set $V$ will be doubly
exponential in the input size (which is the size of the input formula plus the size
of the input GTRS). 
In a second step, we will solve the FO-theory of a simple 
word structure $\Trans^+$ on an alternating Turing machine. More precisely,
we will show the following result:

\begin{thm} \label{thm:string}
There exists an alternating Turing-machine $M$, which accepts precisely those
pairs $( \Trans,\varphi)$, where $\Trans$ is a finite 
labelled graph and $\varphi$ is an FO-sentence over the signature of 
$\Trans$ with $\Trans^+ \models \varphi$.
Moreover, $M$ runs in time 
$O(n^{\ell+1} \cdot |\varphi|)$, where $n$ is the number of nodes of
$\Trans$ and $\ell$ is the quantifier
rank of $\varphi$. Finally, the number of alternations is 
bounded by $O(\ell)$.
\end{thm}
We prove Theorem~\ref{thm:string} in Section~\ref{sec:string}.
Together with our first reduction, Theorem~\ref{thm:string}
yields Theorem~\ref{thm:upper-main}.

\subsection{Proof of Theorem~\ref{thm:upper-main}}

In this section, we will prove Theorem~\ref{thm:upper-main}.
Let $\R = (A,\Sigma,R)$ be a GTRS over the ranked
alphabet $A$ and let $r$ be the maximal size of a 
tree that appears in $R$. Let $\Trans = \Trans(\R)$ and
let $\varphi$ be an FO-sentence of quantifier 
rank $\ell+1$ over the signature of $\Trans$. 
We want to check, whether $\Trans \models \varphi$.
Define the sets
\begin{eqnarray*}
\Ar & = & \{ m \in \N \mid m \geq 1, A_m \neq \emptyset \}, \\
M & = & \{ 1 + \sum_{m \in \Ar} d_m \cdot (m-1) \mid d_m \in \N \text{
  for } m \in \Ar\} .
\end{eqnarray*}
Note that by Lemma~\ref{lemma:number-leaves},
we have $n \in M$ if and only if there exists a tree (or chain) $t \in \T_A$
with exactly $n$ leaves.
Also note that $M = \N \setminus \{0\}$ in case
$A_2 \neq \emptyset$. Let 
$$p = \max(\Ar) \geq 2$$ 
denote the maximal rank of a symbol from $A$. We define a function 
$$
\inter : M \to \N \cup\{\infty\}
$$ 
as follows:
Let $m \in M$. If $A_1 \neq \emptyset$ (i.e., there exists
a unary symbol), then we set $\inter(m) = \infty$.
If $A_1 = \emptyset$, then let $\inter(m)$ be the maximal number of internal nodes in a tree $t \in
\T_A$ with exactly $m$ leaves (this maximum exists if $A_1=\emptyset$; in fact $\inter(m)
\leq m-1$). The intuition behind setting $\inter(m) = \infty$ in case
$A_1 \neq \emptyset$ is that there exist arbitrarily large trees with
$m$ leaves. Note that $\inter(1)=0$.

\begin{lem} \label{lemma-int1}
For every $m \in M$ we have $\inter(m) \geq \frac{m-1}{p-1}$.
\end{lem}

\begin{proof}
It suffices to show the lemma for the case $A_1 = \emptyset$.
In this case, the lemma can be shown by induction on $m$. The case $m = 1$ is clear.
Let $m \in M \setminus \{1\}$ and let $t \in \T_A$ be a tree with $m$
leaves and $\inter(m)$ many internal nodes. 
Let $u \in D_t$ be an internal node such that all children of $u$ are
leaves. Let $t(u) \in A_q$ with $q \geq 2$. If we replace $u$ by a
leaf, we obtain a tree with $\inter(m)-1$ many internal nodes and 
$m-q+1 \in M$ many leaves.
We must have $\inter(m-q+1) =  \inter(m)-1$ (if there would be a tree with
$m-q+1$ leaves and more than $\inter(m)-1$ many internal nodes, then we
would obtain a tree with $m$ leaves and more than $\inter(m)$ many
internal nodes by replacing an arbitrary leaf by a node with $q$
children). Moreover, by induction (note that $q \geq 2$), we 
have $\inter(m-q+1) \geq \frac{m-q}{p-1}$. Hence, we get
$\inter(m) \geq  \frac{m-q}{p-1} + 1 = \frac{m-q+p-1}{p-1} \geq
\frac{m-1}{p-1}$ (since $p \geq q$).
\end{proof}

\begin{lem} \label{lemma-int2}
Assume that $A_1 = \emptyset$.
For every $m \in M$ there exists a chain with $m$ leaves and $\inter(m)$
many internal nodes.
\end{lem}

\begin{proof}
Let $m \in M$.
By definition, there exists a tree $t \in \T_A$ with $m$ leaves and 
$\inter(m)$ internal nodes. It is easy to restructure $t$ into a chain
so that the number of leaves and the number of internal nodes is not 
changed. More precisely, take a tree 
$t = f(t_1,\ldots, t_n)$ (in term
notation) with $m$ leaves and $\inter(m)$ internal nodes, which is not
a chain. By induction, we can assume that every $t_i$ ($1 \leq i \leq n$)
is either a chain or the constant $a \in A_0$ (for some arbitrarily 
chosen $a \in A_0$). Since $t$ is not a chain there exist $1 \leq i <
j \leq n$ such that $t_i$ and $t_j$ are chains.
Choose an arbitrary child $x$ of the the maximal internal node
$\max(t_i)$ of the chain $t_i$; hence $x$ is a leaf of $t_i$. Take the tree
$$
t' = f(t_1,\ldots, t_{i-1}, t_i[x/t_j],t_{i+1},\ldots, t_{j-1},a,t_{j+1},\ldots,t_n).
$$
This tree has the same number of leaves and internal nodes as $t$.
Continuing this way, we finally obtain a chain.
\end{proof}
For numbers $1 \leq i \leq j$ let
$$
T[i,j] = \{t \in \T_A \mid i \leq \size(t) \leq j \}.
$$
For $0 \leq i \leq \ell$ let
\begin{equation} \label{def:sigma}
\sigma(i)  =  \ell \cdot r \cdot 7 \cdot 4^i \cdot 
              \big( (p-1) \cdot r \cdot 4^i + 1 \big) + p \cdot r
              \cdot 4^i \leq r^2 \cdot p \cdot 2^{O(\ell)} .
\end{equation}
Note that we have
\begin{equation} \label{eq-sigma}
\sigma(i+1) \geq \sigma(i) + p \cdot r \cdot 3 \cdot 4^i \geq \sigma(i) + r \cdot 3 \cdot 4^i 
\end{equation}
for all $0 \leq i \leq \ell$.
Let 
$$
U = T[1, \sigma(\ell) + r \cdot p \cdot 4^\ell].
$$
Moreover,  for every $0 \leq i \leq \ell$ let
\begin{eqnarray}
U_i & = & T[1, \sigma(i)] \subseteq U, \nonumber \\
V_i & = & T[1, r \cdot 4^i] \subseteq U, \label{def-V_i} \\
W_i & = & \{ \alpha(u_1,\ldots,u_q) \mid q \geq 1, \alpha \in A_q,
u_1, \ldots, u_q \in V_i\} \setminus V_i \subseteq U. \label{def-W_i}
\end{eqnarray}
Note that $\size(t) \leq r \cdot p \cdot 4^i + 1$ for all $t \in W_i$
and $V_i \cap W_i = \emptyset$.
We consider the set $U$ as a finite alphabet and 
the sets $U_i$, $V_i$, and $W_i$ as subalphabets.
Note that 
\begin{equation} \label{bound-U}
|U| \leq 
|A|^{\sigma(\ell) + r \cdot p \cdot 4^\ell} . 
\end{equation}
Define the language
\begin{equation} \label{def-Z}
Z = \{ w \in U^+ \mid |w| \in M \}
\end{equation}
over the alphabet $U$. Note that $Z = U^+$ in case $A_2 \neq \emptyset$.
On the set $(\N \times Z) \cup U$ we define a labelled graph $\String_1$
with label set $\Sigma$ as follows:
Take an action $\sigma  \in \Sigma$. By our general lifting constructions from
Section~\ref{sec:labelled-graph}, the binary relation $\xrightarrow{\sigma}$
on $\T_A$ is implicitly lifted to a binary relation on
$\T_A^+$ and $\N \times \T_A^+$. Since $(\N \times Z) \cap U = \emptyset$,
$\xrightarrow{\sigma}$ can be viewed as a binary relation 
on $(\N \times Z) \cup U$; simply take the disjoint union of the relations
on $(\N \times Z)$ and $U$. Finally, we define the $\Sigma$-labelled graph
\begin{equation} \label{structure-S1}
\String_1 = ( (\N \times Z) \cup U,\; \Sigma, \; \{ \xrightarrow{\sigma} \mid \sigma \in \Sigma \}).
\end{equation}
For a word $w = u_1 u_2 \cdots u_n \in U^*$ with $u_1, \ldots, u_n \in U$ we define
$$
|\!| w |\!| = \sum_{i=1}^n \size(u_i) . 
$$
We define the sets
\begin{eqnarray}
Z_i & = & \{ w \in  V_i^* W_i V_i^* \cap Z \mid |\!| w |\!| + \inter(|w|) >
\sigma(i) \}, \label{def-Z_i} \\
L_i & = & (\N \times Z_i)  \cup U_i. \nonumber
\end{eqnarray}
Note that $Z_i = V_i^* W_i V_i^* \cap Z$ in case $A_1 \neq \emptyset$
(clearly, we set $n + \infty = \infty$ for every number $n$).
Assume that the first-order sentence $\varphi$ 
is of the form
$Q_\ell x_\ell \;  \cdots \;  Q_1 x_1 \; Q_0 x_0: \psi$, where 
$Q_0, \ldots, Q_\ell \in \{\forall, \exists\}$ and $\psi$
is quantifier-free. For $0 \leq i \leq \ell-1$
and elements $s_{i+1}, \ldots, s_\ell \in (\N \times Z) \cup U$ let
us define the set
$$
L_i(s_{i+1}, \ldots, s_\ell) = L_i \cup S_{3 \cdot 4^i}(\String_1, s_{i+1}, \ldots, s_\ell) .
$$
We define a first-order sentence $\varphi_1$ (with
quantifiers relativized to the sets $L_i(s_{i+1}, \ldots, s_\ell)$)
over the signature of $\String_1$ as 
\begin{equation} \label{def:varphi_1(2)}
\varphi_1 \ = \ Q_\ell x_\ell \in L_\ell \
Q_{\ell-1} x_{\ell-1} \in L_{\ell-1}(x_\ell) \
\cdots \
Q_0 x_0 \in L_0(x_1, \ldots, x_\ell) : \psi.
\end{equation}
We want to show that 
$\Trans \models \varphi$ if and only if $\String_1 \models \varphi_1$.
For this, we need the following lemma, which is the main technical contribution
in this section. The reader might skip the proof at first reading.

\begin{lem} \label{lemma-EFG}
Assume that
\begin{itemize}
\item $0 \leq i \leq \ell$,
\item $\overline{s} = (s_{i+1}, \ldots, s_\ell) \in ((\N \times Z) \cup U)^{\ell-i}$ with
$s_j \in  L_j \cup S_{3 \cdot 4^j}(\String_1, s_{j+1}, \ldots,
s_\ell)$ for all $j \in [i+1, \ell]$,
\item $\overline{t} = (t_{i+1}, \ldots, t_\ell) \in \T_A^{\ell-i}$, and
\item $f: S_{4^{i+1}}(\String_1, \overline{s}) \to S_{4^{i+1}}(\Trans, \overline{t})$
is an isomorphism  such that
$f \rest S_{4^{i+1}}(\String_1, s_j)$ is the identity for 
all $j \in [i+1, \ell]$ with $t_j \in U_{i+1}$ or $s_j \in U_{i+1}$.
\end{itemize}
Then, the following holds:
\begin{enumerate}[label=\({\alph*}]
\item
For all $t_i \in \T_A$ there exists
$s_i \in L_i \cup S_{3 \cdot 4^i}(\String_1,\overline{s})$ 
and an isomorphism 
$$
g : S_{4^i}(\String_1, s_i, \overline{s}) \to 
S_{4^i}(\Trans, t_i, \overline{t})$$ 
such that
$f$ and $g$ are compatible\footnote{Recall the definition of compatible functions from the beginning of Section~\ref{sec:prel}.} and 
$g \rest S_{4^i}(\String_1, s_j)$ is the identity for 
all $j \in [i, \ell]$ with
$t_j \in U_i$ or $s_j \in U_i$.
\item
For all $s_i \in L_i \cup S_{3 \cdot 4^i}(\String_1,\overline{s})$ 
there exists $t_i \in \T_A$ 
and an isomorphism 
$$g : S_{4^i}(\String_1, s_i, \overline{s}) \to 
S_{4^i}(\Trans, t_i, \overline{t})
$$ 
such that
$f$ and $g$ are compatible and 
$g \rest S_{4^i}(\String_1, s_j)$ is the identity for 
all $j \in [i, \ell]$ with
$t_j \in U_i$ or $s_j \in U_i$.
\end{enumerate}
\end{lem}

\noindent Before we prove the lemma, let us provide some intuition.
For case (a) we will basically distinguish two cases:
In case $t_i$ is ``close'' to some tree in the tuple $\overline{t}$, then
the simulating $s_i$ can safely be chosen as $t_i$ itself.
In case $t_i$ is ``far'' to all trees in $\overline{t}$, we distinguish two cases:
Either the size of $t_i$ exceeds $\sigma(i)$ from (\ref{def:sigma}) or not.
If $|t_i|>\sigma(i)$, then $s_i$ will be chosen as a pair from
$\{n\}\times Z_i$ for some fresh number $n$ that does
not appear as a first component of any element in $\overline{s}$, and
where the second component of $s_i$ consists basically 
of $t_i\setminus C$ for some prefix-closed subset $C$ of $t_i$'s nodes.
Intuitively, this means that $s_i$ does not have to be ``too big''
in order to simulate $t_i$: only ``small'' subtrees of $t_i$ have to be accounted
for. Lemma~\ref{lemma-GTRS-spheres} will be crucial.
In case $|t_i|\leq\sigma(i)$, we can prove that we can set
 $s_i=t_i\in U_i$.
For case (b) we can proceed similarly, but the main  
crux is that for each element $s_i\in\N\times Z_i$ we can 
build a tree $t_i\in\T_A$ such that the spheres of radius $4^i$ 
around $s_i$ and $t_i$ are isomorphic. 
For building the latter trees, we have to distinguish the case when $A_1\not=\emptyset$
and the case when $A_1=\emptyset$.

\begin{proof}
Let $f: S_{4^{i+1}}(\String_1, \overline{s}) \to S_{4^{i+1}}(\Trans, \overline{t})$
be an isomorphism such that
$f \rest S_{4^{i+1}}(\String_1, s_j)$ is the identity for 
all $i+1 \leq j \leq \ell$ with
$t_j \in U_{i+1}$ or $s_j \in U_{i+1}$.
Let us first prove statement (a). Let $t_i \in \T_A$.
We distinguish two cases:

\medskip
\noindent
{\em Case 1.} $t_i \in S_{3 \cdot 4^i}(\Trans,\overline{t})$.
Note that this implies that $t_i$ belongs to the range of 
the isomorphism $f$ and that
$$S_{4^i}(\Trans,t_i, \overline{t}) 
\subseteq S_{4^{i+1}}(\Trans, \overline{t}).$$
Then, we set $s_i = f^{-1}(t_i) \in S_{3 \cdot 4^i}(\String_1,\overline{s})$.
We define $g$ as the restriction of $f$ to the set
$S_{4^i}(\String_1, s_i, \overline{s}) \subseteq S_{4^{i+1}}(\String_1, \overline{s})$.
Now, assume that $t_i \in U_i$, i.e., $\size(t_i) \leq \sigma(i)$.
We have to show that $f \rest S_{4^i}(\String_1,s_i)$ is the identity.
Let $t_j$ ($i+1 \leq j \leq \ell$) such that $d_\Trans(t_i,t_j) \leq 3
\cdot 4^i$.
With Lemma~\ref{lemma-dist} it follows 
$$
\size(t_j) \leq  \size(t_i) + r \cdot 3 \cdot 4^i \leq \sigma(i) + r \cdot 3 \cdot 4^i
\stackrel{\text{\eqref{eq-sigma}}}{\leq} \sigma(i+1) .
$$
Hence, $t_j \in U_{i+1}$ and 
$f \rest S_{4^{i+1}}(\String_1, s_j)$ is the identity.
Since  $d_{\String_1}(s_i,s_j) =  d_\Trans(t_i,t_j) \leq 3 \cdot 4^i$, we have
$S_{4^i}(\String_1,s_i) \subseteq S_{4^{i+1}}(\String_1, s_j)$.
It follows that $f \rest S_{4^i}(\String_1,s_i)$
is the identity. If $s_i \in U_i$,
then we can argue analogously.

\medskip
\noindent
{\em Case 2.} $t_i \not\in S_{3 \cdot 4^i}(\Trans,\overline{t})$.
We will find $s_i \in L_i$ and an isomorphism 
$f' : S_{4^i}(\String_1,s_i) \to S_{4^i}(\Trans,t_i)$ such that
$s_i \not\in S_{3 \cdot 4^i}(\String_1,\overline{s})$.  
Then, Lemma~\ref{lemma:disjoint-spheres} implies 
that $g = (f\rest S_{4^i}(\String_1,\overline{s})) \uplus f'$ is an isomorphism from 
$S_{4^i}(\String_1,s_i, \overline{s})$ to 
$S_{4^i}(\Trans,t_i, \overline{t})$, which is compatible with $f$.
Moreover, we will show that if $t_i \in U_i$ or $s_i \in U_i$,
then $f'$ is the identity.

In order to find $s_i$,
let  $t_i \setminus \up(t_i, r \cdot 4^i) =  u_1 \cdots u_m$.
Recall that the latter string is the lexicographic order of all maximal subtrees
of $t_i$ whose size is at most $r\cdot 4^i$. Hence,
$\size(u_j) \leq r \cdot 4^i$ for each $j$, i.e., $u_j \in V_i$
(see \eqref{def-V_i}).

\medskip
\noindent
{\em Case 2.1.} $\size(t_i)  > \sigma(i)$. We must have $t_i \neq
u_1$, because otherwise  $\size(t_i) \leq r\cdot 4^i \leq \sigma(i)$,
which is a contradiction. 
Therefore, there must exist $1 \leq j \leq m$, a  symbol $\alpha \in A$
of rank $q \geq 1$, and a prefix-closed subset $C \subseteq \up(t_i, r \cdot 4^i)$ such
that $\alpha(u_j, \ldots, u_{j+q-1}) \in W_i$ (see \eqref{def-W_i}) and
$$
t_i \setminus C = u_1 \cdots u_{j-1} \alpha(u_j, \ldots, u_{j+q-1})
u_{j+q} \cdots u_m.
$$
Let $w = t_i \setminus C$. 
By Lemma~\ref{lemma:number-leaves(2)}, we have $|w| \in M$.
By the definition of the mapping $\inter$,
we have $|\!| w |\!| + \inter(|w|) \geq \size(t_i)$ and
hence $|\!| w |\!| + \inter(|w|) > \sigma(i)$ by assumption. 
Thus, we get $w \in Z_i$ by definition of $Z_i$ in \eqref{def-Z_i}.
Choose a number $n \in \N$ such that
$n$ does not appear as a first component of a pair from
$\{s_{i+1}, \ldots, s_\ell\}  \cap (\N \times Z)$.
Finally, we set
$$
s_i = (n,  w)
\in \N \times Z_i \subseteq L_i .
$$
Due to the choice of $n$, we have $s_i \not\in S_\rho(\String_1,\overline{s})$ for all $\rho$.
Moreover, with Lemma~\ref{lemma-GTRS-spheres} we get
$S_{4^i}(\String_1,s_i) \cong S_{4^i}(\Trans,t_i)$.
Finally, $\size(t_i) > \sigma(i)$, i.e., $t_i \not\in U_i$, and $s_i \not\in U$.

\medskip
\noindent
{\em Case 2.2.} $\size(t_i) \leq \sigma(i)$, i.e.,
$t_i \in U_i$.
We set  $s_i = t_i \in U_i$.
Note that $S_{4^i}(\Trans,t_i) \subseteq U$, which implies 
$S_{4^i}(\String_1,s_i) = S_{4^i}(\Trans,t_i)$.
Assume that $s_i \in S_{3\cdot 4^i}(\String_1, \overline{s})$. We will deduce a contradiction.
Let $i+1 \leq j \leq \ell$ such that $d_{\String_1}(s_i,s_j) \leq 3 \cdot 4^i$.
Since $s_i \in U$, we must have $s_j \in U$ as well (there is no path
in $\String_1$ between the sets $U$ and $\N \times Z$).
Moreover, with Lemma~\ref{lemma-dist} we get 
$$
\size(s_j) \leq \size(s_i) + r \cdot 3 \cdot 4^i \leq \sigma(i)+ r \cdot 3 \cdot 4^i 
\stackrel{\text{\eqref{eq-sigma}}}{\leq} \sigma(i+1),
$$
i.e., $s_j \in U_{i+1}$. This implies that
$f \rest S_{4^{i+1}}(\String_1, s_j)$ is the identity.
Hence, $t_i \in S_{3\cdot 4^i}(\Trans, t_j)$,
a contradiction. 
We can finally choose for $f'$ 
the identity isomorphism on $S_{4^i}(\String_1,s_i) = S_{4^i}(\Trans,t_i)$.
This proves (a).

Let us now prove (b). Let $s_i \in L_i \cup S_{3 \cdot 4^i}(\String_1,\overline{s})$. Again, we distinguish two cases.

\medskip
\noindent
{\em Case 1.} $s_i \in S_{3 \cdot 4^i}(\String_1,\overline{s})$.
This implies $$S_{4^i}(\String_1,s_i,\overline{s}) 
\subseteq S_{4^{i+1}}(\String_1, \overline{s}).$$
We set $t_i = f(s_i) \in S_{3 \cdot 4^i}(\Trans,\overline{t})$.
We can conclude as in Case 1 for the proof of point (a) above.

\medskip
\noindent
{\em Case 2.} $s_i \not\in S_{3 \cdot 4^i}(\String_1,\overline{s})$.
Hence, $s_i \in L_i$.
We will find $t_i \in \T_A$ and an isomorphism 
$f' : S_{4^i}(\String_1,s_i) \to S_{4^i}(\Trans,t_i)$ 
such that
$t_i \not\in S_{3 \cdot 4^i}(\Trans, \overline{t})$.
Then, Lemma~\ref{lemma:disjoint-spheres} implies 
that the mapping $g = (f\rest S_{4^i}(\String_1,\overline{s})) \uplus f'$ is an isomorphism from 
$S_{4^i}(\String_1,s_i, \overline{s})$ to 
$S_{4^i}(\Trans,t_i, \overline{t})$, which is compatible with $f$.
Moreover, we will show that if $t_i \in U_i$ or $s_i \in U_i$,
then $f'$ is the identity.

\medskip
\noindent
{\em Case 2.1.} $s_i \in U_i \subseteq \T_A$. 
We set $t_i = s_i \in U_i$,
which implies $S_{4^i}(\Trans,t_i) \subseteq U$.
Thus, $S_{4^i}(\String_1,s_i) = S_{4^i}(\Trans,t_i)$.
Assume that $t_i \in S_{3 \cdot 4^i}(\Trans, \overline{t})$. We will deduce a contradiction.
Let $i+1 \leq j \leq \ell$ such that $d_{\Trans}(t_i,t_j) \leq 3 \cdot 4^i$.
Lemma~\ref{lemma-dist} implies 
$$
\size(t_j) \leq \size(t_i) + r \cdot 3 \cdot 4^i \leq \sigma(i)+ r \cdot 3 \cdot 4^i 
\stackrel{\text{\eqref{eq-sigma}}}{\leq} \sigma(i+1).
$$
This implies that
$f \rest S_{4^{i+1}}(\String_1, s_j)$ is the identity.
Hence, $s_i \in S_{3\cdot 4^i}(\String_1, s_j)$,
a contradiction. 
We can finally choose for $f'$ 
the identity isomorphism on $S_{4^i}(\String_1,s_i) = S_{4^i}(\Trans,t_i)$.

\medskip
\noindent
{\em Case 2.2.} $s_i \in \N \times Z_i$.
Let $s_i = (n, u_1 \cdots u_m)$ with $u_1, \ldots, u_m \in V_i \cup
W_i$, $m \in M$, and $|\!| u_1 \cdots u_m |\!| + \inter(m) > \sigma(i)$.
There is exactly one $1 \leq j \leq m$ 
with $u_j \in W_i$. Let $u_j = \alpha(v_1,\ldots, v_q)$ with 
$q \geq 1$, $\alpha \in A_q$, and
$v_1, \ldots, v_q \in V_i$.
Define the string
\begin{equation}\label{def:string-w}
w = u_1 \cdots u_{j-1} v_1 \cdots v_q u_{j+1} \cdots u_m 
\end{equation}
of length $m+q-1$. Since $m\in M$, we also have $m+q-1\in M$.

\medskip
\noindent
{\em Case 2.2.1.} $A_1 \neq \emptyset$.
Then, we can
choose for $t_i$ a tree with the following properties:
\begin{itemize}
\item $t_i \setminus \up(t_i, r\cdot 4^i) = w$. For this, we connect all trees $u_1,
  \ldots, u_m$ to one tree using a chain of symbols of rank at least
  2, starting from $u_j \in W_i$. Since $m \in M$, this is possible by 
 Lemma~\ref{lemma-int2} (applied to the ranked alphabet $A \setminus A_1$).
\item $t_i \not\in S_{3 \cdot 4^i}(\Trans,\overline{t})$ and
  $\size(t_i) > \sigma(i)$. This can be
  enforced by adding a long enough chain of unary symbols to the root.
\end{itemize}
With Lemma~\ref{lemma-GTRS-spheres}, the first point implies $S_{4^i}(\String_1,s_i) \cong
S_{4^i}(\Trans,t_i)$. Moreover, since $\size(t_i) > \sigma(i)$, we have $t_i \not\in U_i$.

\medskip
\noindent
{\em Case 2.2.2.} $A_1 = \emptyset$ and thus $\inter(m) < \infty$.
Note that $|\!| w|\!| = |\!| u_1 \cdots u_m |\!|-1$, i.e.,
$$
|\!| w|\!| + \inter(m)  = |\!| u_1 \cdots u_m |\!| + \inter(m) - 1 \geq \sigma(i) .
$$
Every tree in the string $w$ has size at most 
$r  \cdot 4^i$.  
Hence, we have $|\!| w|\!| \leq (m+q-1) \cdot r\cdot 4^i$.
We get
$$
(m+q-1) \cdot r\cdot 4^i + \inter(m) \geq \sigma(i).
$$
Moreover, since  $\inter(m) \geq \frac{m-1}{p-1}$ by Lemma~\ref{lemma-int1}, we have
$m+q-1 \leq \inter(m) \cdot (p-1) + q \leq \inter(m) \cdot (p-1) + p$.
We get
$$
(\inter(m) \cdot (p-1) + p) \cdot r\cdot 4^i + \inter(m) \geq \sigma(i).
$$
Solving this inequality for $\inter(m)$ yields
$$
\inter(m) \geq \frac{\sigma(i) - p \cdot r \cdot 4^i}{ (p-1) \cdot r
  \cdot 4^i + 1}.
$$
Plugging in the definition of $\sigma(i)$ from \eqref{def:sigma} yields
\begin{equation}\label{lower-bound-lambda}
\inter(m) \geq \ell \cdot r \cdot 7 \cdot 4^i .
\end{equation}
We now define $\ell+1$ different trees $t'_1, \ldots, t'_{\ell+1}$
as follows.

We first fix a sequence $\alpha_1, \ldots, \alpha_{\inter(m)}$
of symbols from $A \setminus A_0$
such that that every chain, where the $j^{th}$ internal
node is labelled with $\alpha_j$ has exactly $m$
leaves. By Lemma~\ref{lemma-int2} such a sequence exists.
In the following, we consider chains with $\inter(m)+1$ many
internal nodes such that the following hold:
\begin{itemize}
\item The $j^{\text{th}}$ internal node ($1 \leq j \leq \inter(m)$)
is labelled with $\alpha_j$ and the maximal internal node is 
labelled with $\alpha \in A_q$ (thus, such a chain has $m+q-1$ leaves). 
\item Every internal node belongs to $\{1,2\}^*$ (thus, every internal
  node, which is not the root, is either the first or the second child
  of its parent node).
\item All leaves in the chain are
labelled with some fixed constant $\Box \in A_0$.
\end{itemize}
This means that such a chain is uniquely determined by its
maximal internal node $u = \max(t) \in \{1,2\}^{\inter(m)}$. 
We write $t = \chain(u)$. 

Let $u, v \in \{1,2\}^{\inter(m)}$ such that $u = xay$ and $v = xbz$
with $x,y,z \in \{1,2\}^*$, $a,b \in \{1,2\}$, $a \neq b$.
Define $\diff(u,v) = |y|+1$ ($= |z|+1$). Recall also the definition
of the $\diff$-value for two trees from Section~\ref{sec:trees}.
Then, we have
\begin{equation}\label{ineq:diff}
\diff(\chain(u), \chain(v)) > \diff(u,v).
\end{equation}
In fact, $\diff(\chain(u), \chain(v)) \geq 2 \cdot \diff(u,v)$ holds.

Since $\inter(m) \geq \ell \cdot
r \cdot  7 \cdot 4^i$ by \eqref{lower-bound-lambda}, 
we can find $\ell+1$ strings
$w_1, \ldots, w_{\ell+1} \in \{1,2\}^{\inter(m)}$ such that 
for all $k \neq k'$ we have
\begin{equation} \label{lower-bound-diff}
\diff(w_k, w_{k'}) \geq r \cdot  7 \cdot4^i .
\end{equation}
We may for instance set
$$
w_k = 1^{\inter(m) - \ell \cdot r \cdot  7 \cdot4^i} 1^{(k-1) \cdot r \cdot
  7 \cdot4^i} 2^{(\ell-k+1) \cdot r \cdot
  7 \cdot4^i} .
$$
Let us define the chain $c_k = \chain(w_k)$ for all $1 \leq k \leq
\ell+1$. Hence, \eqref{ineq:diff} and \eqref{lower-bound-diff} imply
\begin{equation}\label{ineq:c_k-diff}
\diff(c_k, c_{k'}) > r \cdot  7 \cdot4^i 
\end{equation}
for all $k \neq k'$.
Moreover, every chain $c_k$ has exactly $m + q -1$ leaves.
Finally, the tree $t'_k$ is obtained
from the chain $c_k$ as follows: We replace 
the $q$ children of the maximal internal node $\max(c_k)$ 
(which is labelled with $\alpha \in A_q$) by 
$v_1, \ldots, v_q$ (in this order). All other $m-1$ leaves 
are replaced by the trees  $u_1, \ldots, u_{j-1}, u_{j+1}, \ldots, u_m$
(the order does not matter). 
It follows that the string $t'_k \setminus \up(t'_k, r \cdot 4^i)$
is a permutation of the string $w$ from 
\eqref{def:string-w}.
With Lemma~\ref{lemma-GTRS-spheres} and Remark~\ref{rem-GTRS-spheres}
this ensures that
$$
S_{4^i}(\String_1,s_i) \cong S_{4^i}(\Trans,t_k')
$$
for all $1 \leq k \leq \ell+1$.
Moreover, since each of the trees $u_1, \ldots, u_{j-1}, 
v_1, \ldots, v_q, u_{j+1}, \ldots, u_m \in V_i$ 
has size at most $r  \cdot 4^i$,
the number of nodes in the subtree of $t_k'$ rooted at a leaf of $c_k$
may grow by at most $r  \cdot 4^i$, when we replace the 
leaf by one of the trees 
$u_1, \ldots, u_{j-1}, v_1, \ldots, v_q$, $u_{j+1}, \ldots, u_m$.
This implies
$$
\diff(t_{k}', t_{k'}') \geq \diff(c_k, c_{k'}) - r  \cdot 4^i
\stackrel{\text{\eqref{ineq:c_k-diff}}}{>} r \cdot  6 \cdot 4^i,
$$
provided $k \neq k'$.
Hence, Lemma~\ref{lemma-diff} implies
\begin{equation} \label{eq-dist>44^j}
d_{\Trans}(t_k', t_{k'}') > 6 \cdot 4^i 
\end{equation}
for all $k \neq k'$.
We claim that there is at least one 
$1 \leq k \leq \ell+1$ such that 
$t_k' \not\in S_{3 \cdot 4^i}(\Trans,\overline{t})$.
In order to obtain a contradiction, assume that
for each $1 \leq k \leq \ell+1$ there exists some $t_h$
($i+1 \leq h \leq \ell$) such that 
$d_{\Trans}(t_k', t_h) \leq  3 \cdot 4^i$.
Since there are only $\ell-i \leq \ell$ such trees $t_h$,
the pigeon hole principle implies that there exist $k \neq k'$ and $h$ with
$d_{\Trans}(t_k', t_h) \leq  3 \cdot 4^i$
and $d_{\Trans}(t_{k'}', t_h) \leq  3 \cdot 4^i$.
Hence, $d_{\Trans}(t_{k}',t_{k'}') \leq 6 \cdot 4^i$, which
contradicts \eqref{eq-dist>44^j}.
We finally set $t_i = t_k'$, where $k$ is chosen such that
$t_k' \not\in S_{3 \cdot 4^i}(\Trans,\overline{t})$.
Finally, note that $\size(t_i) = |\!| u_1 \cdots u_m |\!| + \inter(m) > \sigma(i)$
(i.e., $t_i \not\in U_i$) and $s_i \not\in U$.
This concludes the proof of the lemma.
\end{proof}
Lemma~\ref {lemma-EFG} allows us to prove
the following lemma:

\begin{lem} \label{lemma-EFG2}
Assume that
\begin{itemize}
\item $-1 \leq i \leq \ell$,
\item $\overline{s} = (s_{i+1}, \ldots, s_\ell) \in ((\N \times Z) \cup U)^{\ell-i}$ with
$s_j \in  L_j \cup S_{3 \cdot 4^j}(s_{j+1}, \ldots, s_\ell)$ for all
$j \in [i+1, \ell]$,
\item $\overline{t} = (t_{i+1}, \ldots, t_\ell) \in \T_A^{\ell-i}$, and
\item $f: S_{4^{i+1}}(\String_1, \overline{s}) \to S_{4^{i+1}}(\Trans, \overline{t})$
is an isomorphism  such that
$f \rest S_{4^{i+1}}(\String_1, s_j)$ is the identity for 
all $j \in [i+1, \ell]$ with $t_j \in U_{i+1}$ or $s_j \in U_{i+1}$.
\end{itemize}
Then, for every quantifier-free first-order formula $\psi$ over the signature of $\Trans$ and 
all quantifiers $Q_0, \ldots, Q_i \in \{\forall,\exists\}$ we have
\begin{gather*}
\String_1 \models Q_i x_i  \in L_i(\overline{s}) \cdots Q_0 x_0 \in
L_0(x_1, \ldots, x_i,\overline{s}) : \psi(x_0, \ldots, x_i, \overline{s}) \\
\quad \Longleftrightarrow \quad \\
\Trans \models Q_i x_i  \cdots Q_0 x_0  : \psi(x_0, \ldots, x_i, \overline{t}) .
\end{gather*}
\end{lem}

\begin{proof}
The lemma can be shown by induction on $i$, starting with $i = -1$. 
For the induction base ($i=-1$), note that the existence of the isomorphism
$f$ ensures that $\overline{s}$ and $\overline{t}$ satisfy the same quantifier-free
formulas.
The induction step uses Lemma~\ref{lemma-EFG} and the 
classical back-and-forth argument from
the proof of the Ehrenfeucht-Fra\"iss\'{e}-Theorem.
\end{proof}
Setting $i=\ell$ in Lemma~\ref{lemma-EFG2}, it follows 
$\Trans \models \varphi$ if and only if $\String_1 \models
\varphi_1$, where $\varphi_1$ is from \eqref{def:varphi_1(2)}. 

For the remainder of the poof of Theorem 4.1, we proceed as follows:
We simplify the sentence $\varphi_1$ (which is not an ordinary first-order sentence due
to the additional constraints for the variables $x_0, \ldots, x_\ell$)
and the structure $\String_1$ further, so that
we can finally apply Theorem~\ref{thm:string}. 
In a first step (Step 1 below), we eliminate in the formula $\varphi_1$ the relativation 
of the variables $x_i$ to the spheres $S_{3 \cdot 4^i}(\String_1, x_{i+1}, \ldots, x_\ell)
\subseteq L_i(x_{i+1}, \ldots, x_\ell)$. Then, the structure $\String_1$ will be 
freplaced by an isomorphic structure $\String_3$ (using an intermediate isomorphic
copy $\String_2$).  These is done in Step 2 and Step 3 below.
The structure $\String_3$ will be almost of the form 
$\mathfrak{T}^+$ for a finite labelled graph $\mathfrak{T}$ (these are the structures 
appearing in Theorem~\ref{thm:string}). The only difference is that the universe
of $\String_3$ is a regular language of the form $\Delta^* \Theta \Delta^*$ 
(for finite alphabets $\Delta$ and $\Theta$) instead of the set of all non-empty finite
words (as it is the case for $\mathfrak{T}^+$). Also the constraint sets $L_i\subseteq L_i(x_{i+1}, \ldots, x_\ell)$ from $\String_1$
will be mapped to simple regular languages in $\String_3$. We finally transform 
$\String_3$ into a structure $\String_4 = \mathfrak{T}^+$ by enlarging the finite alphabet
over which words from $\String_3$ are defined (Step 4).

\medskip
\noindent
{\em Step 1.}
Recall that quantifiers in $\varphi_1$ are relativized
to the sets $$L_i(x_{i+1}, \ldots, x_\ell) = L_i \cup 
S_{3 \cdot 4^i}(\String_1, x_{i+1}, \ldots, x_\ell).$$
Note that $x_i \in S_{3 \cdot 4^i}(\String_1, x_{i+1}, \ldots,
x_\ell)$ means that
$\bigvee_{j = i+1}^\ell d_{\String_1}(x_i,x_j) \leq 3 \cdot 4^i$
holds. By Lemma~\ref{lemma:Fischer-Rabin} we can find 
an equivalent first-order formula of size 
$O((\ell-i) \cdot i + (\ell-i) \cdot |\Sigma|)
\leq O(\ell^2 + \ell \cdot |\Sigma|)$
and quantifier rank $O(i) \leq O(\ell)$ 
(we take the formula 
$\theta(x,y) = (x=y \vee \bigvee_{\sigma\in \Sigma} \sigma(x,y) \vee
\sigma(y,x))$ in Lemma~\ref{lemma:Fischer-Rabin}; note that
the binary representation of $3  \cdot 4^i$ has only 2 1-bits).
After replacing the constraints $x_i \in S_{3 \cdot 4^i}(\String_1, x_{i+1}, \ldots,
x_\ell)$ for $1 \leq i \leq \ell$, the resulting equivalent sentence has size 
$|\varphi| +  O(\ell^3 + \ell^2 \cdot |\Sigma|)$ and 
quantifier rank $O(\ell)$.

\medskip
\noindent
{\em Step 2.}
It remains to eliminate constraints of the form 
$x_i \in L_i = (\N \times Z_i) \cup U_i$. 
In order to do this, we will
change the  labelled graph $\String_1$ to a labelled graph
of the form $\mathfrak{T}^+$ for a finite labelled graph $\mathfrak{T}$. 
The basic idea will be to change the alphbet $U$ by taking words
over of $U$ of some bounded length as the new symbols; the resulting
alphabet will be the set $U' \cup U''$ below. 

In the following, we assume that $A_1 = \emptyset$ (and hence
$\inter(m) < \infty$ for all $m$); the case
$A_1 \neq \emptyset$ is the simpler one.

In order to cope with the length constraint $|w| \in M$ in the
definition of the set $Z_i$ from \eqref{def-Z_i}, 
we define for $0 \leq i \leq \ell$ the sets
\begin{eqnarray*}
U' & = & \{ w \in U^+ \mid |w|+1 \in \Ar \},  \\
V'_i & = & \{ w \in V_i^+ \mid |w|+1 \in \Ar \} \subseteq U'.
\end{eqnarray*}
We have 
\begin{equation} \label{bound-U'}
|U'| \leq (|U|+1)^{p-1} \stackrel{\text{\eqref{bound-U}}}{\leq} 
(|A|+1)^{(p-1) \cdot (\sigma(\ell) + p \cdot r \cdot 4^\ell)}
\stackrel{\text{\eqref{def:sigma}}}{\leq} 
|A|^{2^{O(\ell)} p^2 \cdot r^2} .
\end{equation}
Moreover, for $0 \leq i \leq \ell$ let us define
$W'_i$ as the set of all minimal words (with respect to the factor relation on words)
$w \in V_i^* W_i V_i^*$ with $|w| \in M$ (and hence $w \in Z$ by \eqref{def-Z}) and
$|\!| w |\!| + \inter(|w|) > \sigma(i)$ (and hence $w \in Z_i$ by \eqref{def-Z_i}).
It follows that for such a word $w$ we have
$$
|\!| w |\!| + \inter(|w|) - (p-1) \cdot r \cdot 4^i - 1 \leq \sigma(i) .
$$
Since $|w| \leq |\!| w |\!|$ and $\inter(|w|) \geq \frac{|w|-1}{p-1}$
by Lemma~\ref{lemma-int1}, we have
$$
|w| + \frac{|w|-1}{p-1} - (p-1) \cdot r \cdot 4^i - 1 \leq \sigma(i)
$$
or equivalently
$$
|w| \leq \frac{p-1}{p} \cdot \sigma(i) + \frac{(p-1)^2}{p} \cdot r
\cdot 4^i + 1 .
$$
Hence, for all $w \in W'_i$ we have 
\begin{equation}\label{length-bound-w}
|w| \leq \sigma(i) + p\cdot r\cdot 4^i + 1 .
\end{equation}
Let us set 
\begin{equation}\label{def-gamma}
\gamma = \sigma(\ell) + p\cdot r\cdot 4^\ell + 1
\stackrel{\text{\eqref{def:sigma}}}{\leq} 
p \cdot r^2 \cdot 2^{O(\ell)} + p\cdot r\cdot 4^\ell + 1
= p \cdot r^2 \cdot 2^{O(\ell)},
\end{equation}
which is an upper bound for the right-hand side of 
\eqref{length-bound-w}. Note that $\gamma$ is exponential in our input size.
Let
$$
U''  =  \{ w \in Z \mid |w| \leq  \gamma \},
$$
which contains all alphabets $W'_i$ ($0 \leq i \leq \ell$) as well as $U$.
We have 
\begin{equation} \label{bound-U''}
|U''| \leq (|U|+1)^{\gamma}
\stackrel{\text{\eqref{bound-U}}}{\leq} 
(|A|+1)^{\gamma \cdot (\sigma(\ell) + p \cdot r \cdot 4^\ell)}
\stackrel{\text{\eqref{def:sigma}}}{\leq} 
|A|^{2^{O(\ell)} p^2 \cdot r^4}
\end{equation}
which is doubly exponential in our input size.

For the further discussion, it is important that elements of $U' \cup U''$ are viewed
as single symbols. For a word $w \in ({U'} \cup {U''})^*$ we can define an expanded word
$\mathsf{exp}(w) \in U^*$ in the natural way (e.g. $\mathsf{exp}((a) (abba) (b)
(ba)) = aabbabba$). 
Note that for every word $w \in {U'}^* U'' {U'}^*$ we have
$|\mathsf{exp}(w)| \in M$ (i.e., $\mathsf{exp}(w) \in Z$). 
Vice versa, for every word  $w \in Z$ there exists at least one word 
$w' \in {U'}^* U'' {U'}^*$ with $\mathsf{exp}(w') = w$.
Moreover, for every word $w \in {V_i'}^* W'_i {V_i'}^*$ we have
$|\mathsf{exp}(w)| \in M$  and 
$|\!| \mathsf{exp}(w) |\!| + \inter(|\mathsf{exp}(w)|) > \sigma(i)$.
Vice versa, if $w \in V_i^* W_i V_i^* \cap Z$ with 
$|\!| w |\!| + \inter(|w|) > \sigma(i)$ (i.e., $w \in Z_i$), then 
there exists at least one word $w' \in {V_i'}^* W'_i {V_i'}^*$
with $\mathsf{exp}(w') = w$.
This allows us to replace the constraint set
$Z_i = \{ w \in  V_i^* W_i V_i^* \cap Z \mid |\!| w |\!| + \inter(|w|) > \sigma(i)\}$ by the 
set ${V_i'}^* W'_i {V_i'}^*$.
Note that for a word $w \in Z_i$ there may exist several 
words $w' \in {V_i'}^* W'_i {V_i'}^*$ with $\mathsf{exp}(w') = w$.
This is not a problem: by taking the set $\N \times Z_i$ 
in the structure $\String_1$, we basically take  $\aleph_0$
many copies of $w$.

By our lifting
construction from 
Section~\ref{sec:labelled-graph}, every binary
relation 
$\xrightarrow{\sigma}$ ($\sigma \in \Sigma$)
on $\T_A$ is defined on $U' \cup U'' \subseteq \T_A^+$
and hence on $(\N \times {U'}^* U'' {U'}^*) \cup U$.
Using this, it follows that our   labelled graph
$\String_1 = ( (\N \times Z) \cup U,\; \Sigma, \; \{ \xrightarrow{\sigma} \mid \sigma \in \Sigma \})$
is isomorphic to the  labelled graph
$$\String_2 =  ( (\N \times {U'}^* U'' {U'}^*) \cup U,\; \Sigma, \; \{ \xrightarrow{\sigma} \mid \sigma \in \Sigma \}).$$
The isomorphism maps the constraint set 
$L_i = (\N \times Z_i) \cup U_i$ to $(\N \times {V_i'}^* W'_i {V_i'}^*) \cup U_i$.

\medskip
\noindent
{\em Step 3.}
In order to get rid of the direct product with $\N$ in  $\N \times {V_i'}^* W'_i {V_i'}^*$
we add a new symbol $\$$ to the alphabet $U' \cup U''$. 
We lift the relations $\xrightarrow{\sigma}$ ($\sigma \in \Sigma$)
from ${U''}^*$ to $(U'' \cup \{\$\})^*$ in the standard way ($\$$ 
does not occur in the left-hand and right-hand sides of the relations
$\xrightarrow{\sigma}$). 
Then, the labelled graph $\String_2$ 
(and hence $\String_1$) is isomorphic to 
the graph
$$
\String_3 =  ( (U' \cup \{\$\})^* U'' (U' \cup \{\$\})^* \cup U, \; \Sigma, \; \{ \xrightarrow{\sigma} \mid \sigma \in \Sigma \}).
$$
The isomorphism maps $U$ identically to $U$ and the set $\N \times \{w\}$ (for 
$w \in {U'}^* U'' {U'}^*$) is mapped bijectively onto 
the set of those words from $(U' \cup \{\$\})^* U'' (U' \cup \{\$\})^*
\setminus U$, whose projection 
onto the subalphabet $U''$ is $w$. 
Hence, the constraint set 
$\N \times {V_i'}^* W'_i {V_i'}^*$ is mapped to the set
\begin{equation} \label{constraint set}
(V'_i \cup \{\$\})^* W'_i (V'_i \cup \{\$\})^* \setminus U.
\end{equation}
{\em Step 4.}
In order to express in first-order logic that a word belongs to the above constaint set
\eqref{constraint set},
we introduce another symbol $\#$. Hence, our final alphabet is 
$$
\Gamma = U' \cup U'' \cup \{\$, \# \} .
$$
With \eqref{bound-U'} and \eqref{bound-U''},
the size of $\Gamma$ can be estimated as
\begin{equation} \label{size-Gamma}
|\Gamma| = 2+|U'|+|U''|
\leq |A|^{2^{O(\ell)} p^2 \cdot r^4}.
\end{equation}
Next, we define 
a finite labelled graph
$\mathfrak{T} = (\Gamma, \Sigma', \{ \xrightarrow{a} \mid a \in
\Sigma' \})$ 
with node set $\Gamma$ as follows.
The set of actions is
$$
\Sigma' = \Sigma \cup \Gamma.
$$
The set of transitions is defined as follows. By our lifting
construction from 
Section~\ref{sec:labelled-graph}, every binary
relation 
$\xrightarrow{\sigma}$ ($\sigma \in \Sigma$)
on $\T_A$ is defined on $\Gamma$ ($\$$ and $\#$ 
do not occur in the left-hand and right-hand sides of the relations
$\xrightarrow{\sigma}$).
Moreover, for  $a \in \Gamma$ we define the relation
$$
\xrightarrow{a} \ \; = \ \{ (a,\#) \}.
$$
Finally, using the construction from Section~\ref{sec:labelled-graph},
we define the labelled graph
$$
\String_4 =  \mathfrak{T}^+ .
$$
We will construct a sentence $\varphi_4$ over the signature
of $\String_4$ such that $\String_1 \models \varphi_1$ if
and only if $\String_4 \models \varphi_4$.
Using the edge relations $\xrightarrow{a}$ 
($a \in \Gamma$), we can express $x \in \Omega^+$ (for $\Omega
\subseteq \Gamma$) as 
$$
\bigwedge_{a \in \Gamma \setminus \Omega} \neg\exists y : a(x,y) .
$$ 
Moreover, a constraint $|x|_{\Omega} \geq k$ 
(saying that there are at least $k$ occurrences of symbols from
$\Omega$ in the word $x$)
can be expressed as
$$
\exists y_1, \ldots, y_k \bigg( \bigwedge_{j \neq j'} y_j \neq y_{j'}
\wedge \bigwedge_{j \in [1,k]} \bigvee_{a \in \Omega} a(x,y_i) \bigg) .
$$
This allows us to express e.g. $|x|_\Omega = k$ or $x \in \Omega$.
Hence, a constraint $x \in L_i = ( \N \times Z_i)  
\cup U_i$ in $\varphi_1$ can be
replaced by the formula
$$
 \big( x \in (V'_i \cup W'_i \cup \{ \$ \})^+ \ \wedge \
|x|_{W'_i} = 1 \ \wedge \ x \not\in U \big) \
\vee \ x \in U_i
$$
of size $O(|\Gamma|)$.
(for the correctness of this formula it is important that
$V'_i \cap W'_i = \emptyset$ which follows from $V_i \cap W_i = \emptyset$).
The size of the resulting sentence $\varphi_4$
can be bounded by 
$|\varphi| + O(\ell^3 + \ell^2 \cdot |\Sigma| + \ell \cdot |\Gamma|)$ and 
its quantifier rank is still $O(\ell)$.

\medskip
\noindent
We can now conclude the proof of Theorem~\ref{thm:upper-main}.
Recall that our overall goal is to check, whether $\Trans \models \varphi$ holds.
By the above constructions, this is equivalent to $\String_4 \models \varphi_4$.
By Theorem~\ref{thm:string}, this can be decided on an alternating Turing machine in time 
$$
O(|\Gamma|^{O(\ell)} \cdot |\varphi_4|)
\leq \poly( |\Gamma|^{O(\ell)} + |\varphi| + |\Sigma|) 
$$
using $O(\ell) \leq O(|\varphi|)$ many alternations.
Recall from \eqref{size-Gamma} that $|\Gamma| \leq |A|^{2^{O(\ell)} p^2 \cdot r^4}$.
Hence, we can bound the running time by
$\poly( |A|^{2^{O(\ell)} p^2 \cdot r^4} + |\varphi| + |\Sigma|)$,
which is doubly exponential in the input size. This concludes
the proof of Theorem~\ref{thm:upper-main}.

\subsection{Proof of Theorem~\ref{thm:string}} \label{sec:string}

Let us fix a finite labelled graph $\mathfrak{G} = (V, \Sigma,
\{\xrightarrow{\sigma} \mid \sigma  \in \Sigma\})$ and let $n = |V|$.
We want to decide the first-order theory of $\mathfrak{G}^+$. For this
we can w.l.o.g.  assume that $n \geq 2$.
Moreover, we can assume that $\Sigma = V \times V$ and that 
the edge from $a \in V$ to $b \in V$ is labelled with $(a,b)$ (the original edge
relations are definable by disjunctions in this new graph).
Our decision procedure for the first-order theory of $\mathfrak{G}^+$ uses
the method of Ferrante and Rackoff from Section~\ref{sec:FO} for the function
$H(k,\ell) = n^{k+\ell+2}+k$. For this, we define a suitable 
equivalence relation $\equiv_{k,\ell}$ on $k$-tuples over $V^*$. 
The definition of this equivalence relation uses a simpler equivalence
relation $\equiv_d$ defined on words, which corresponds to counting 
and comparing symbols up to the threshold $d$. The main combinatorial lemma
for the equivalence $\equiv_{k,\ell}$ is Lemma~\ref{lemma:stringEF1}. It rougly
says that if $\overline{u} \equiv_{k,\ell} \overline{v}$ and $u \in V^*$, then one can
always find a ``short'' word $v$ such that $(\overline{u},u) \equiv_{k,\ell} (\overline{v},v)$.
This corresponds to point (b) in Theorem~\ref{thm-FR}. 
To apply the method of Ferrante and Rackoff, we also have to show that 
 $\overline{u} \equiv_{k,0} \overline{v}$ implies that  $\overline{u}$ and  $\overline{v}$
 satisfy the same quantifer-free formulas in $\mathfrak{G}^+$
 (point (a) in Theorem~\ref{thm-FR}). This is stated in Lemma~\ref{lemma:stringEF2}.

Recall that for a word $u \in A^*$ over a finite alphabet $A$ 
and $a \in A$,
$|u|_a$ denotes the number of occurrences of $a$ in $u$. 
For $d \geq 1$ and $u,v \in A^*$, we write 
$u \equiv_d v$ if for all 
$a \in A$ the following holds:
\begin{itemize}
\item  $|u|_a = |v|_a$ or 
\item  ($|u|_a \geq d$ and $|v|_a \geq d$)
\end{itemize}
Note that $\equiv_d$ is an equivalence relation and
that $u \equiv_{d+1} v$ implies $u \equiv_d v$.

Let $A$ and $B$ be finite alphabets.
For two words $u = a_1 a_2 \cdots a_k \in A^*$ and $v = b_1 b_2
\cdots b_k \in B^*$ of the same length $k$ we define
the convolution 
$$u \otimes v = (a_1,b_1) (a_2,b_2) \cdots (a_k,b_k) 
\in (A \times B)^k.
$$

\begin{lem}\label{lemma-comb}
Let $\alpha \in \N$, $u,v \in \Gamma^*$ (where $\Gamma$ is a finite alphabet), 
$u' \in V^*$ with $|u|=|u'|$, $u \equiv_{\alpha\cdot n} v$,
and $|v| \geq \alpha \cdot n \cdot |\Gamma|$.
Then there exists $v' \in V^*$ with 
$|v|=|v'|$  and $u \otimes u'  \equiv_\alpha v \otimes v'$.
\end{lem}

\begin{proof}
Let $a \in \Gamma$ and $b \in V$.
Consider the values $m_{a,b} = |u \otimes u'|_{(a,b)}$
and $n_a = |v|_a$. 
Finding a word $v' \in V^*$ such that $|v'| = |v|$ 
and $u \otimes u'  \equiv_\alpha v \otimes v'$
is equivalent to finding
numbers $n_{a,b}$ (which will be $|v \otimes v'|_{(a,b)}$)
such that 
\begin{itemize}
\item
$\sum_{b \in V} n_{a,b} = n_a$ for all $a \in \Gamma$ and 
\item 
$m_{a,b} = n_{a,b}$ or ($m_{a,b} \geq \alpha$ and $n_{a,b} \geq
\alpha$) 
for all $a\in \Gamma$, $b \in V$.
\end{itemize}
Note that $u \equiv_{\alpha \cdot n} v$
implies
$$
\sum_{b \in V} m_{a,b} = n_a \ \text{ or } \ 
(\sum_{b \in V} m_{a,b} \geq \alpha\cdot n \text{ and } n_a \geq
\alpha\cdot n) 
$$
for all $a \in \Gamma$.
Also recall that $|V|=n$. 
We choose the numbers $n_{a,b}$ as follows, where $a \in \Gamma$:
\begin{itemize}
\item 
If $\sum_{b \in V} m_{a,b} = n_a$, then 
we set $n_{a,b} = m_{a,b}$ for all $b \in V$.
\item 
If $\sum_{b \in V} m_{a,b} \geq \alpha\cdot n$
and $n_a \geq \alpha\cdot n$, then
(since $|V|=n$) there must be 
at least one $b \in V$ with $m_{a,b} \geq \alpha$.
We first set $n_{a,b} = m_{a,b}$ for all $b \in V$ with
$m_{a,b} < \alpha$. For all remaining 
$b \in V$ (which satisfy $m_{a,b} \geq \alpha$) we set
$n_{a,b}$ to some value $\geq \alpha$ such that
the total sum $\sum_{b \in V} n_{a,b}$ becomes 
$n_a$. Since $n_a \geq \alpha\cdot n$ this is possible.
\end{itemize}
\end{proof}
For all $k,\ell \in \mathbb{N}$ we define an equivalence relation
$\equiv_{k,\ell}$ on the set $(V^*)^k$ of $k$-tuples of words over $V$
as follows:
Let $(u_1, \ldots, u_k), (v_1, \ldots,v_k) \in (V^*)^k$.
Then $(u_1, \ldots, u_k) \equiv_{k,\ell} (v_1, \ldots,v_k)$ if and
only if the following conditions hold:
\begin{enumerate}[label=\({\alph*}]
\item For all $1 \leq i,j \leq k$, $|u_i| = |u_j|$  if and only if 
  $|v_i| = |v_j|$.
\item For all $1 \leq i \leq k$, $u_i = v_i$ or
$|u_i| \geq n^{k+\ell+1}$ and $|v_i| \geq n^{k+\ell+1}$.
\item For all $1 \leq i \leq k$ the following holds:
Let $1 \leq i_1 < i_2 \cdots < i_m \leq k$ 
be exactly those indices such that $|u_i| = |u_{i_1}| = \cdots
= |u_{i_m}|$. Hence,  $|v_i| = |v_{i_1}| = \cdots
= |v_{i_m}|$ due to (a). Then 
$u_{i_1} \otimes u_{i_2}  \otimes\cdots \otimes u_{i_m} \equiv_\alpha
 v_{i_1} \otimes v_{i_2}  \otimes\cdots \otimes v_{i_m}$, where
$\alpha = n^{\ell+1}$.
\end{enumerate}

\begin{lem} \label{lemma:stringEF1}
Let $k \geq 0$, $\ell > 0$, $(u_1, \ldots, u_k) \equiv_{k,\ell} (v_1, \ldots,v_k)$,
and let  $u_{k+1} \in  V^*$. Then there exists 
 $v_{k+1}  \in  V^*$ such that
 $|v_{k+1}| \leq n^{k+\ell+1}+k$ and
 $(u_1, \ldots, u_k,u_{k+1}) \equiv_{k+1,\ell-1} (v_1, \ldots,v_k,v_{k+1})$.
\end{lem}

\begin{proof}
Assume that $(u_1, \ldots, u_k) \equiv_{k,\ell} (v_1, \ldots,v_k)$
and let
$u_{k+1} \in  V^*$.
We distinguish several cases:

\medskip
\noindent
{\em Case 1.} $|u_{k+1}| \neq |u_i|$ 
for all $1 \leq i \leq k$.

\medskip
\noindent
{\em Case 1.1.}  $|u_{k+1}|<n^{k+\ell+1}$.
Then, we must have $|u_{k+1}| \neq |v_i|$ for all $1 \leq i \leq k$
(if $|v_i| = |u_{k+1}| <n^{k+\ell+1}$, then we must have $u_i = v_i$ 
by (b) and hence $|u_{k+1}| = |u_i|$).
We set $v_{k+1} = u_{k+1}$.

\medskip
\noindent
{\em Case 1.2.}  $|u_{k+1}|\geq n^{k+\ell+1}$.
Choose a number $\lambda$ with $n^{k+\ell+1} \leq \lambda \leq
n^{k+\ell+1}+k$ and $|v_i| \neq \lambda$ 
for all $1 \leq i \leq k$.
We will find a word $v_{k+1}$ such that
$|v_{k+1} | = \lambda$ and  
$u_{k+1} \equiv_\alpha v_{k+1}$ for 
$\alpha = n^{\ell}$.
Since $|u_{k+1}|\geq n^{k+\ell+1}$, there 
exists a symbol $a \in V$ such that
$|u_{k+1}|_a \geq n^{k+\ell} \geq n^{\ell} = \alpha$.
If $\lambda \geq |u_{k+1}|$, then we simply increase
the number of occurrences of $a$ in $u_{k+1}$
until a word of length $\lambda$ is reached.
If $\lambda < |u_{k+1}|$, then $|u_{k+1}| > n^{\ell+1}$.
Hence, there even exists
$a \in V$ with $|u_{k+1}|_a > n^{\ell}$.
We remove one of the occurrences of $a$ in $u_{k+1}$.
We can repeat this step until a word of length 
$\lambda$ is reached.

\medskip
\noindent
{\em Case 2.} $|u_{k+1}| = |u_i|$ 
for some $1 \leq i \leq k$. Let $1 \leq i_1 < i_2 \cdots < i_m \leq k$ 
be exactly those indices such that $|u_{k+1}| = |u_{i_1}| = \cdots =
|u_{i_m}|$. Let $u = u_{i_1} \otimes u_{i_2}  \otimes\cdots \otimes
u_{i_m}$ and $v = v_{i_1} \otimes v_{i_2}  \otimes\cdots \otimes
v_{i_m}$. Point (c) implies $u \equiv_{n^{\ell+1}} v$.

\medskip
\noindent
{\em Case 2.1.}  $|u_{k+1}| < n^{k+\ell+1}$.
Hence, we have $|u_i| < n^{k+\ell+1}$.
This implies  $|u_i| = |v_i| < n^{k+\ell+1}$ by (b).
We set $u_{k+1} = v_{k+1}$.
Note that $v_{i_1} = u_{i_1}, \ldots, v_{i_m} = u_{i_m}$ by (b). This implies
$$u_{i_1} \otimes u_{i_2}  \otimes\cdots \otimes u_{i_m} \otimes u_{k+1} \equiv_\alpha
 v_{i_1} \otimes v_{i_2}  \otimes\cdots \otimes v_{i_m}\otimes v_{k+1}$$
for  all $\alpha$.

\medskip
\noindent
{\em Case 2.2.}  $|u_{k+1}| \geq n^{k+\ell+1}$.
Hence, we have $|u| = |u_i| \geq n^{k+\ell+1}$.
This implies  $|v| = |v_i| \geq n^{k+\ell+1}$ by (b).
We have to choose a word $v_{k+1}$ with $|v_{k+1}| = |v_i|$ and 
$u \otimes u_{k+1} \equiv_\alpha v \otimes v_{k+1}$
for  $\alpha = n^{\ell}$. 
This is possible by Lemma~\ref{lemma-comb}:
Note that $\alpha \cdot n = n^{\ell+1}$
and thus $u \equiv_{\alpha \cdot n} v$.
In order to apply Lemma~\ref{lemma-comb} we set in addition
$u' = u_{k+1}$, $v' = v_{k+1}$, and $\Gamma = V^m$.
This implies 
$$|v| \geq n^{k+\ell+1} = n^{\ell} \cdot n \cdot
n^k  \geq n^{\ell} \cdot n \cdot n^m = \alpha \cdot n \cdot |\Gamma|.
$$
Hence, Lemma~\ref{lemma-comb} can be applied indeed.
\end{proof}
Recall the definition of the infinite graph $\mathfrak{G}^+$
from \eqref{def-G^+}.

\begin{lem} \label{lemma:stringEF2}
If $(u_1, \ldots, u_k) \equiv_{k,0} (v_1, \ldots,v_k)$,
then the tuples $(u_1, \ldots, u_k)$ and $(v_1, \ldots,v_k)$
satisfy the same quantifier-free formulas in the 
graph $\mathfrak{G}^+$.
\end{lem}

\begin{proof}
By symmetry, it suffices to prove the following two points:
\begin{enumerate}[label=\({\alph*}]
\item If $u_i = u_j$ then also $v_i = v_j$.
\item If $u_i \xrightarrow{(a,b)} u_j$ for some $(a,b) \in V \times V$ then
also $v_i \xrightarrow{(a,b)} v_j$.
\end{enumerate}
Let us first prove (a). W.l.o.g. assume that $i=1$ and $j=2$.
Let $2 < i_1 < i_2 < \cdots < i_m$ be those indices such that
$|u_1| = |u_2| = |u_{i_1}| = \cdots = |u_{i_m}|$.
Since $(u_1, \ldots, u_k) \equiv_{k,0} (v_1, \ldots,v_k)$, we get
$|v_1| = |v_2| = |v_{i_1}| = \cdots = |v_{i_m}|$
and $u_1 \otimes u_2 \otimes u_{i_1} \otimes \cdots \otimes u_{i_m}
\equiv_\alpha v_1 \otimes v_2 \otimes v_{i_1} \otimes \cdots \otimes
v_{i_m}$ for $\alpha = n \geq 2$. Since $u_1 = u_2$, all symbols that occur in 
$u_1 \otimes u_2 \otimes u_{i_1} \otimes \cdots \otimes u_{i_m}$
are of the form $(a,a,\cdots)$ for some $a \in V$.
Hence, the same has to hold for $v_1 \otimes v_2 \otimes v_{i_1} \otimes \cdots \otimes
v_{i_m}$. But this means that $v_1 = v_2$.

For point (b), assume first that $a=b$. Thus, $u_i = u_j$
and $|u_i|_a > 0$. By point (a), we already know that $v_i = v_j$.
If $i=j$, then we can w.l.o.g. assume that $i=j=1$.
Let $1 < i_1 < i_2 < \cdots < i_m$ be those indices such that
$|u_1| = |u_{i_1}| = \cdots = |u_{i_m}|$.
Since $(u_1, \ldots, u_k) \equiv_{k,0} (v_1, \ldots,v_k)$, we get
$|v_1| = |v_{i_1}| = \cdots = |v_{i_m}|$
and $u_1 \otimes u_{i_1} \otimes \cdots \otimes u_{i_m}
\equiv_\alpha v_1 \otimes v_{i_1} \otimes \cdots \otimes
v_{i_m}$ for $\alpha = n \geq 2$.
Since $|u_1|_a > 0$, the word $u_1 \otimes u_{i_1} \otimes \cdots
\otimes u_{i_m}$ contains at least one occurrence of a symbol 
of the form $(a,\ldots)$. Hence, the same holds for
$v_1 \otimes v_{i_1} \otimes \cdots \otimes
v_{i_m}$. But this means that $v_1 \xrightarrow{(a,a)} v_1$.
If $i \neq j$, then we can argue similarly. 

Finally, let us assume that $a \neq b$. We must have $i \neq j$.
W.l.o.g. assume that $i=1$ and $j=2$. Let
us choose the indices $2 < i_1 < i_2 < \cdots < i_m$ as for the proof
of  point (a) above. Since $u_1 \xrightarrow{(a,b)} u_2$,
the following holds for the word 
$u = u_1 \otimes u_2 \otimes u_{i_1} \otimes \cdots \otimes u_{i_m}$:
$u$ contains exactly one occurrence of a symbol of the form
$(a,b,\ldots)$ and all other symbols in $u$ are of the form 
$(c,c,\ldots)$ for $c \in V$. Again, the same has
to be true for $v_1 \otimes v_2 \otimes v_{i_1} \otimes \cdots \otimes
v_{i_m}$ (only here it is important that $n \geq 2$ and not just $n
\geq 1$). Hence, $v_1 \xrightarrow{(a,b)} v_2$.
\end{proof}
We can now prove Theorem~\ref{thm:string}.
Let $\varphi = Q_0 x_0 \cdots Q_\ell x_\ell : \psi(x_0,\ldots,x_\ell)$
be a first-order formula of quantifier rank $\ell+1$
over the signature of 
$\mathfrak{G}^+$, where $Q_0, \ldots, Q_\ell \in \{\forall,\exists\}$
and $\psi$ is quantifier-free.
For $0 \leq i \leq \ell$ 
let $L_i = \{ w \in V^+ \mid |w| \leq n^{\ell+2}+i\}$.
Theorem~\ref{thm-FR}
(with $H(k,\ell) = n^{k+\ell+2}+k$),
Lemma~\ref{lemma:stringEF1}, and \ref{lemma:stringEF2}
imply that $\mathfrak{G}^+ \models \varphi$ if and only
if 
$$
\mathfrak{G}^+ \models Q_1 x_0 \in L_0 \cdots Q_\ell x_\ell \in L_\ell : \psi(x_0,\ldots,x_\ell).
$$
This can be decided on an alternating Turing machine in time 
$O(n^{\ell+2} \cdot |\varphi|)$
with $\ell$ alternations by guessing words $v_i \in L_i$ 
either existentially (if $Q_i = \exists$) or universally (if $Q_i =
\forall$) and then verifying the statement $\psi(x_0,\ldots,x_\ell)$.

\section{An $\ATIME(2^{2^{\poly(n)}},\poly(n))$  lower bound}

In this section, we will prove that there exists a fixed GTRS
such that the corresponding ground tree rewrite graph has 
an $\ATIME(2^{2^{\poly(n)}},\poly(n))$-complete first-order theory.
This will be achieved using a suitable tiling problem. 
Tiling problems turned out to be an important tool for proving
hardness and undecidability results in logic, see e.g. \cite{BoGrGu01}.
In a first step we will prove hardness for $2\NEXP$ (doubly
exponential non-deterministic time) in Section~\ref{S Lower}.
In Section~\ref{section:pushing}, we will finally push the lower
bound to $\ATIME(2^{2^{\poly(n)}},\poly(n))$.

\subsection{Tiling systems}

A {\em tiling system} is a tuple $S=(\Theta,\Ho,\Ve)$, where $\Theta$ is a finite set of 
{\em tile types}, $\Ho\subseteq \Theta\times\Theta$ is a {\em horizontal matching
relation}, and $\Ve\subseteq\Theta\times\Theta$ is a {\em vertical matching
relation}.
A mapping $\sigma:[0,k-1]\times[0,k-1]\rightarrow\Theta$ (where $k\geq 0$)
is a {\em $k$-solution for $S$} if 
for all $(x,y)\in[0,k-1]\times[0,k-1]$ the following holds:
\begin{itemize}
\item if $x<k-1$,  $\sigma(x,y)=\theta$, and $\sigma(x+1,y)=\theta'$, then 
$(\theta,\theta')\in \Ho$, and
\item if $y<k-1$, $\sigma(x,y)=\theta$, and $\sigma(x,y+1)=\theta'$, then 
$(\theta,\theta')\in \Ve$.
\end{itemize}
Let $\Sol_k(S)$ denote the set of all $k$-solutions for $S$.
Let $w=w_0\cdots w_{n-1} \in\Theta^n$ be a word and let $k\geq n$.
With $\Sol_k(S,w)$ we denote the set of all 
$\sigma \in \Sol_k(S)$ such that 
$\sigma(x,0)=w_x$ for all $x\in[0,n-1]$.
For a tiling system $S$ we define its
{\em $(2^{2^n}\times 2^{2^n})$ tiling problem} as follows:\\[-0.4cm]

\bigskip

\problemx{$(2^{2^n}\times 2^{2^n})$ tiling problem for tiling system
$S=(\Theta,\Ho,\Ve)$}
{A word $w\in\Theta^n$.}
{Does $\Sol_{2^{2^n}}(S,w) \neq \emptyset$ hold?}

\bigskip

\noindent
The following proposition is folklore, see also \cite{BoGrGu01,Chlebus84}.

\begin{prop}{\cite{BoGrGu01,Chlebus84}}{\label{P NEXP}}
There is some fixed tiling system $S_0$ whose $(2^{2^n}\times 2^{2^n})$ tiling problem is
$2\NEXP$-hard under logspace reductions.
\end{prop}

\subsection{Hardness for $2\NEXP$} \label{S Lower}

\newcommand{\ol}{\overline}
\renewcommand{\hat}{\widehat}

Let us fix the tiling system $S_0=(\Theta_0,\Ho_0,\Ve_0)$ of 
Proposition \ref{P NEXP} whose tiling
problem is hard for $2\NEXP$.
We now define a fixed GTRS $\R_0=(A,\Sigma,R)$ and prove that
the first-order theory of $\Trans(\R_0)$ is
$2\NEXP$-hard under logspace reductions.
We define 
\begin{eqnarray*}
A_0&=&\{\heartsuit,\O,\O_\dagger,\O_\ddagger,\Z,
\Z_\dagger,\Z_\ddagger\},\\
A_1&=&\Theta_0,\\
A_2&=&\{\bullet\},\text{ and}\\
\Sigma&\quad=\quad&\{\ell,r,h,u,m_\dagger,m_{\ddagger},
\}\cup\Theta_0\cup A_0.\\
\end{eqnarray*}
The set of rewrite rules $R$ is given as follows:
\begin{enumerate}
\item $X\hook{X}X$ for each $X\in A_0$,
\item $X\hook{m_{\dagger}}X_\dagger$ for each $X\in\{\O,\Z\}$ (this will correspond
to {\em marking} a leaf),
\item $X_\dagger\hook{m_{\ddagger}}X_\ddagger$ for each $X\in\{\O,\Z\}$
(this will correspond to {\em selecting} a leaf),
\item $X_\dagger\hook{h}\heartsuit$ for each $X\in\{\O,\Z\}$,
\item $\bullet(\heartsuit,\heartsuit)\hook{u}\heartsuit$, 
\item $\theta(X_{\ddagger})\hook{\theta}\theta(X_{\ddagger})$ for all
  $\theta\in\Theta_0$, $X\in\{\O,\Z\}$,
\item $\bullet(\heartsuit,X_{\ddagger})\hook{r}X_{\ddagger}$ 
for each $X\in\{\O,\Z\}$, and
\item
$\bullet(X_{\ddagger},\heartsuit)\hook{\ell}X_{\ddagger}$ 
for each $X\in\{\O,\Z\}$.
\end{enumerate}
For the rest of this section we fix $\Trans_0=\Trans(\R_0)$.
Let us fix an input $w=\theta_0\cdots\theta_{n-1}\in\Theta^n$ of the $(2^{2^n}\times
2^{2^n})$ tiling problem for $S_0$.
Our goal is to compute in logspace from $w$ a first-order sentence $\varphi$ 
over $\Sigma$ such that
$$
\Sol_{2^{2^n}}(S_0,w) \neq \emptyset \quad\Longleftrightarrow\quad
\Trans_0\models\varphi.
$$
For each subset $\Gamma\subseteq\Sigma$, we define
$\trans{\Gamma}\ =  \bigcup_{\gamma\in\Gamma}\trans{\gamma}$.
The following lemma follows immediately from
Lemma~\ref{lemma:Fischer-Rabin}
(take the formula $\theta(x,y) = \bigvee_{\gamma \in \Gamma} \gamma(x,y)$).

\begin{lem}{\label{L Fischer}}
Given a subset of actions $\Gamma\subseteq\Sigma$ and $j\in[0,2^{n+1}]$ (in
binary) one can compute in
logspace a first-order formula $\Gamma^j(x,y)$ such that for all
$t,t'\in\T_A$ we have $\Trans_0 \models\Gamma^j(t,t')$ if and only if
$t \; (\trans{\Gamma})^j \;  t'$ in $\Trans_0$.
\end{lem}

In case $\Gamma=\{\gamma\}$ is a singleton, we also write $\gamma^j(x,y)$ for the
formula $\Gamma^j(x,y)$ of Lemma \ref{L Fischer}.
Moreover, for subsets $\Gamma_1, \ldots, \Gamma_k \subseteq \Sigma$
and $j_1, \ldots, j_k \in \N$, we write
$[\Gamma_1^{j_1} \cdots \Gamma_k^{j_k}](x,y)$ for the formula
$$
\exists x_0, \ldots, x_k : \big(x_0 = x \wedge x_k = y \wedge
\bigwedge_{i=1}^k \Gamma_i^{j_i}(x_{i-1},x_i)\big).
$$
A tree $t\in\T_A$ is a {\em tile tree} if $t=\theta(t')$ for some 
$t'\in\T_A$ such that the following holds:
\begin{itemize}
\item $\theta\in\Theta_0$,
\item The label of every leaf of $t'$ is from $\{\Z,\O\}$.
\item The distance of every leaf of $t'$ to the root of $t'$ is $n+1$.
\item Every internal node of $t'$ is labeled with $\bullet$.
\end{itemize}

\begin{exa} \label{ex-tile-tree}
This is a tile tree in case $n+1=3$:
$$
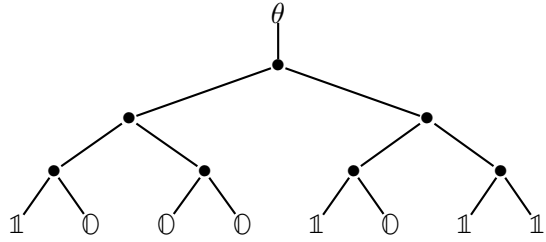
\pstree[levelsep=20pt]
{\Tr{\theta}}
{\pstree[levelsep=20pt]
{\Tr{\bullet}}
{
\pstree[levelsep=20pt]
{\Tr{\bullet}}
{ 
\pstree[levelsep=20pt]
{\Tr{\bullet}}
{\Tr{\O}
\Tr{\Z}
}
\pstree[levelsep=20pt]
{\Tr{\bullet}}
{\Tr{\Z}
\Tr{\Z}
}
}
\pstree[levelsep=20pt]
{\Tr{\bullet}}
{ 
\pstree[levelsep=20pt]
{\Tr{\bullet}}
{\Tr{\O}
\Tr{\Z}
}
\pstree[levelsep=20pt]
{\Tr{\bullet}}
{\Tr{\O}
\Tr{\O}
}
}
}
}
$$
\end{exa}\medskip

Let us fix a tile tree $t$. Note that $t$ has precisely $2^{n+1}=2\cdot 2^{n}$ leaves.
Hence, there is a one-to-one correspondence between $[0,2^{n+1}-1]$ and leaves of
$t$ by means of their lexicographic order in $t$. For each leaf $\lambda$ let
$\lex(\lambda)\in[0,2^{n+1}-1]$ be the position of $\lambda$ among all
leaves w.r.t. the lexicographic order (starting with $0$). 
The intention is that $t$ represents the $\theta$-labeled grid 
element $(M,N)\in[0,2^{2^{n}}-1]^2$, where each leaf $\lambda$ that is a left (resp. right) child
represents the $\lfloor\frac{\lex(\lambda)}{2}\rfloor^{\text{th}}$ least
significant bit of the
$2^n$-bit binary presentation of
$M$ (resp. of $N$): In case $\lambda$ is a left child, then $t(\lambda)=\Z$
 (resp. $t(\lambda)=\O$) if and
only if the $\lfloor\frac{\lex(\lambda)}{2}\rfloor^{\text{th}}$ least
significant bit of $M$ equals $0$ (resp. $1$)
and analogously if $\lambda$ is a right child this corresponds to $N$.
For the tile tree $t$ from Example~\ref{ex-tile-tree} we have
$M = 1+4+8=13$ and $N=8$.

We say a leaf $\lambda$ of a tree $t$ is {\em marked} 
if $t(\lambda)=X_\dagger$ for some $X\in\{\Z,\O\}$.
We say a leaf $\lambda$ of a tree $t$ is {\em selected} 
if $t(\lambda)=X_\ddagger$ for some $X\in\{\Z,\O\}$.
A {\em marked tile tree} is a tree that can be obtained from a tile tree $t$ by marking
{\em every} leaf of $t$.
For the rest of this section, let $D=2^{n+1}-(n+2)$.

\begin{lem}
One can compute in logspace a first-order formula $\marked(x)$ 
such that for every tree $t\in\T_{A \setminus \{\Z_\ddagger,
  \O_\ddagger,\heartsuit \}}$ with
precisely $2^{n+1}$ marked leaves we have:
$\Trans_0 \models\marked(t)$ if and only if the marked leaves of $t$ are
the leaves of some (unique) marked tile subtree of $t$.
\end{lem}

\begin{proof}
The idea is to express the following: Whenever we select any
 of the $2^{n+1}$ marked leaves, we can execute from the resulting tree some sequence
from the language
$$
h^{2^{n+1}-1}u^D\{\ell,r\}^{n+1}\Theta_0.
$$
Let us explain the intuition behind this. 
Assume we have selected exactly one of the $2^{n+1}$ marked leaves of
$t$, and let $t'$ be the resulting tree.
First, note that after executing the sequence
$h^{2^{n+1}-1}$ from $t'$, we have replaced each of the marked leaves
of $t'$ with the symbol $\heartsuit$, reaching some tree $t''$.
Second, when executing $u^D$ from $t''$ we have reached, in case $t$ contained a marked
tile subtree, some tree $t'''$ that has a chain of the following form 
as a subtree, where $X\in\{\Z,\O\}$ and $\theta\in\Theta_0$:
$$
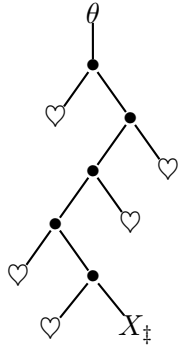
\pstree[levelsep=20pt]
{\Tr{\theta}}
{\pstree[levelsep=20pt]
{\Tr{\bullet}}
{\Tr{\heartsuit}
\pstree[levelsep=20pt]
{\Tr{\bullet}}
{
\pstree[levelsep=20pt]
{\Tr{\bullet}}
{
\pstree[levelsep=20pt]
{\Tr{\bullet}}
{
\Tr{\heartsuit}
\pstree[levelsep=20pt]
{\Tr{\bullet}}
{
\Tr{\heartsuit}
\Tr{X_\ddagger}
}
}
\Tr{\heartsuit}
}
\Tr{\heartsuit}
}
}
}
$$
Finally, from $t'''$ we can now  ``shrink'' this subtree to the tree $\theta(X_\ddagger)$
by executing some sequence from $\{\ell,r\}^{n+1}$ followed by
executing $\theta$.
Formally, we define $\marked(x)$ as follows:
$$
\forall y\left( m_\ddagger(x,y) \rightarrow
\exists z : [h^{2^{n+1}-1} u^D \{\ell,r\}^{n+1}\Theta_0](y,z) \right)
$$
Note that in this formula, $y$ runs over all trees that can be
obtained by selecting a marked leaf of $x$. Basically, in this way
we quantify over all marked leaves of $x$. Note that the formula 
$\marked(x)$ ensures that the marked leaves of $x$ are all at the 
same depth in $x$.
\end{proof}
A {\em grid tree} is a tree $t$ for which every leaf is inside
a subtree of $t$ that is a tile tree.

\newcommand{\grid}{\mathsf{grid}}
\begin{lem}
One can compute in logspace a first-order formula $\grid(x)$ such that
for all $t\in\T_A$ we have $\Trans_0\models\grid(t)$ if and only if
$t$ is a grid tree.
\end{lem}

\begin{proof}
The formula $\grid$ will be a conjunction of the following two statements:
(i) every leaf is either labeled with $\Z$ or $\O$, 
(ii) for each leaf of $t$ that we can mark via the action $m_\dagger$, 
we can mark $2^{n+1}-1$ further
leaves 
reaching some tree $t'$ with $\Trans_0\models\marked(t')$.
Formally, $\grid(x)$ is the conjunction of 
$$\bigwedge_{a\in A_0\setminus\{\Z,\O\}} \neg a(x,x),$$
which realizes (i), and the formula 
$$\forall y \left( m_\dagger(x,y) \rightarrow
\exists z
\big(m_\dagger^{2^{n+1}-1}(y,z)\wedge\marked(z)\big)\right),$$
which realizes (ii). 
\end{proof}
A {\em marked grid tree} is a tree that can be obtained from a grid tree $t$ by
replacing exactly one tile subtree of $t$ by some marked tile tree.
A {\em selected grid tree} is a tree that can be obtained from some marked grid
tree $t$ by selecting {\em precisely one} marked leaf $\lambda$ of
$t$. In that case, $\lex(\lambda) \in [0,2^{n+1}-1]$ is the
lexicographical position of $\lambda$ within the marked tile tree.

\begin{lem}
One can compute in logspace for each $i\in[1,n+1]$ a first-order formula
$\bit_i(x)$ such that for every selected grid tree $t$ with selected
leaf $\lambda$ we have that
the $i^{\text{th}}$ least significant bit of $\lex(\lambda)$ is $1$ if and only
if $\Trans_0\models\bit_i(t)$.
\end{lem}

\begin{proof}
We define $\bit_i(x) \ = \ 
\exists y : [h^{2^{n+1}-1} u^D \{\ell,r\}^{i-1} r](x,y)$.
\end{proof}

\begin{lem}
One can compute for each $\circ\in\{<,=\}$
in logspace a first-order formula $\varphi_{\circ}(x,y)$ such that for
every two selected grid trees $t_1$ and $t_2$ with selected leaves $\lambda_1$
and $\lambda_2$ we have $\Trans_0\models\varphi_{\circ}(t_1,t_2)$ if and only if
$\lex(\lambda_1)\circ\lex(\lambda_2)$.
\end{lem}
\proof
We only treat the case when $\circ$ equals $<$; its definition 
should be self-explanatory:
$$
\bigvee_{j\in[1,n+1]}\bigg(
\left( \neg\bit_j(x)\wedge\bit_j(y)\right)
\wedge
\bigwedge_{1\leq i<j}\left(
\bit_i(x)\leftrightarrow\bit_i(y)\right)
\bigg)\eqno{\qEd}
$$\medskip

\newcommand{\lef}{\mathsf{left}}
\noindent Recall that the unique marked tile subtree of a marked grid
tree $t$ represents a
$\theta$-labeled grid element
$(M,N)\in[0,2^{2^n}-1]^2$ for some $\theta\in\Theta_0$.
Therefore, let us define $M(t)=M$, $N(t)=N$, and $\Theta_0(t)=\theta$.

\begin{lem}
One can compute in logspace FO-formulas $\varphi_{\theta}(x)$,
$\varphi_{i,M}(x,x')$, 
$\varphi_{i,N}(x,x')$, where $\theta\in\Theta_0$ and $i\in\{0,1\}$
such that for all marked grid trees $t$ and $t'$ the following holds:
\begin{enumerate}
\item $\Trans_0\models\varphi_{\theta}(t)$ if and only if $\Theta_0(t)=\theta$,
\item $\Trans_0\models\varphi_{i,M}(t,t')$ if and only if 
$M(t)+i= M(t')$, and
\item $\Trans_0\models\varphi_{i,N}(t,t')$ if and only if $N(t)+i= N(t')$. 
\end{enumerate}
\end{lem}
\proof
For point (1) we define $\varphi_\theta(x)$ as follows:
$$
\exists y : [m_\ddagger h^{2^{n+1}-1} u^D \{\ell,r\}^{n+1} \theta](x,y)
$$
For the remaining points (2) and (3), we only give the formula
$\varphi_{1,M}(x,x')$, i.e., we wish to express that
for any two marked grid trees $t$ and $t'$ we have $\Trans_0\models\varphi_{1,M}(t,t')$
if and only if $M(t)+1=M(t')$. 
Let us fix two marked grid trees $t$ and $t'$. Assume we have selected among the
$2^{n+1}$ marked leaves of $t$ some leaf $\lambda$.
Recall that $\lambda$ represents one of the $2^{n}$ bit positions of $M(t)$ if
and only if $\lambda$ is a {\em left} child, otherwise it would represent a bit
position of $N(t)$. 
Hence we will only be interested in leaves of $t$ and $t'$ which are left
children. For this sake, let us express that the selected leaf of a
selected grid tree $z$ is a left child via the formula 
$\lef(z)$:
$$
\lef(z)\quad=\quad\exists z',z''\left( h(z,z') \wedge
\ell(z',z'') \right)
$$
Our formula $\varphi_{1,M}(x,y)$ is defined as follows:
$$
\exists x', y' \left( 
m_\ddagger(x,x') \wedge m_\ddagger(y,y') \wedge
\varphi_{=}(x',y')\wedge \Z_\ddagger(x',x') \wedge 
\O_\ddagger(y',y')  \wedge
\lef(x') \wedge \psi_1\wedge\psi_2\right) .
$$
Thus, we select a position $p \in [0,2^n-1]$ that is set to $0$ (resp. $1$) in
the binary representation of $M(t)$ (resp. $M(t')$).
The formula $\psi_1(x,y,x',y')$ is defined as
\begin{align*}
& \forall z \bigg( \big( m_\ddagger(x,z) \wedge \varphi_{<}(z,x')\wedge
\lef(z)\big) \rightarrow \O_\ddagger(z,z) \bigg) \wedge \\ 
& \forall z \bigg( \big( m_\ddagger(y,z) \wedge \varphi_{<}(z,y')\wedge
\lef(z)\big) \rightarrow \Z_\ddagger(z,z) \bigg) .
\end{align*}
It expresses that each bit at some position that is smaller than
$p$ is set to $1$ (resp. $0$) in the binary representation of $M(t)$
(resp. $M(t')$). The formula $\psi_2$ 
expresses that the binary representations of 
$M(t)$ and $M(t')$ agree on each position that is bigger than $p$.
Formally, $\psi_2(x,y,x',y')$ is defined as
$$
\forall u, v\bigg(\left( m_\ddagger(x,u) \wedge m_\ddagger(y,v) \wedge
\varphi_{=}(u,v)\wedge\varphi_{<}(x',u)\wedge\lef(u)\right)\rightarrow
 ( \O_\ddagger(u,u) \leftrightarrow\ 
\O_\ddagger(v,v))  \bigg).\eqno{\qEd}
$$

We define the FO-formula $\sol(x)$ as the conjunction of the following
formulas, where $\marky(z_1,z_2)$ is an abbreviation for
$m_{\dagger}^{2^{n+1}}(z_1,z_2)\wedge \marked(z_2)$:
\begin{itemize}
\item $x$ is a grid tree:
$$
\grid(x)
$$
\item 
Whenever we mark two tile subtrees of $x$ that represent the same grid element,
their $\Theta$-labels agree:
$$
\forall y, z \bigg( \left(\marky(x,y)\wedge\marky(x,z)
\wedge\varphi_{0,M}(y,z)\wedge\varphi_{0,N}(y,z)
\right)
\rightarrow
\bigwedge_{\theta\in\Theta_0}\!\!\left(\varphi_\theta(y)\leftrightarrow\varphi_\theta(z)\right)\bigg)
$$
\item Whenever we mark a tile subtree of $x$ that corresponds to the grid
element $(M,N)$ and $M<2^{2^n}-1$ there exists some tile subtree of $x$ that
corresponds to the grid element $(M+1,N)$ and the horizontal matching relation
is satisfied:
\begin{align*}
\forall y \bigg( & \big(\marky(x,y)\wedge
\exists z (m_\ddagger(y,z) \wedge
\Z_\ddagger(z,z) \wedge\lef(z))\big)\rightarrow\\
& \exists y'\big(
\marky(x,y')
\wedge\varphi_{1,M}(y,y')\wedge\varphi_{0,N}(y,y')\wedge
\bigvee_{(\theta,\theta')\in\Ho_0} (\varphi_\theta(y)\wedge\varphi_{\theta'}(y'))
\big) \bigg)
\end{align*}
\item Analogously to the previous formula, we can express that
whenever we mark a tile subtree of $x$ that corresponds to the grid
element $(M,N)$ and $N<2^{2^n}-1$ there exists some tile subtree of $x$ that
corresponds to the grid element $(M,N+1)$ and the vertical matching relation
is satisfied.
\end{itemize}
Finally we can construct a formula $\varphi_w(x)$ that guarantees that
 grid element $(j,0)$ is labeled by $\theta_j$ (recall that
$w=\theta_0\cdots \theta_{n-1}$) for each $j\in[0,n-1]$:
\begin{align*}
\exists y_0,\ldots, y_{n-1} \bigg(
& 
\bigwedge_{j\in[0,n-1]} (\marky(x,y_j)\wedge\varphi_{\theta_{j}}(y_j))
\ \wedge\ 
\forall z ( m_\ddagger(y_0,z) \rightarrow \Z_\ddagger(z,z))
\ \wedge \\
& \bigwedge_{j\in[1,n-1]}
(\varphi_{1,M}(y_{j-1},y_j)\wedge\varphi_{0,N}(y_{j-1},y_j))
\bigg)
\end{align*}
Our final formula $\varphi$ is defined as
$\varphi = \exists x( \sol(x)\wedge\varphi_w(x))$.
It follows by construction that 
$$
\Sol_{2^{2^n}}(S_0,w) \neq \emptyset \quad \Longleftrightarrow \quad
\Trans_0\models\varphi.
$$
With Proposition~\ref{P NEXP} we get:

\begin{thm}
The first-order theory of $\Trans_0$ is $2\NEXP$-hard under logspace
reductions.
\end{thm}

\subsection{Pushing hardness to $\ATIME(2^{2^{\poly(n)}},\poly(n))$}
\label{section:pushing}

Let us fix a tiling system 
$$S=(\Theta,\Ho,\Ve).
$$
Given $\sigma,\sigma' \in\Sol_k(S)$ we say $\sigma'$ {\em extends} $\sigma$
{\em vertically} if
 $\sigma'(x,0)=\sigma(x,k-1)$ for each $x\in[0,k-1]$. 
Let $\Sol_k(S,\sigma)$ be the set 
of all $\sigma' \in \Sol_k(S)$ such that $\sigma'$ extends $\sigma$
vertically.
The standard encoding of Turing machine computations into tilings
shows that there is a fixed tiling system $S_1=(\Theta_1,\Ho_1,\Ve_1)$ such that
the following problem is hard for $\ATIME(2^{2^{\poly(n)}},\poly(n))$ under
logspace reductions.

\bigskip

\problemx{Linearly alternating $(2^{2^{n}}\times 2^{2^n})$ tiling problem (for
$S_1$)}
{A word $w=\theta_0\theta_1\cdots\theta_{n-1}\in\Theta_1^n$, where $n$ is odd.}
{Does $\exists \sigma_1\in\Sol_{2^{2^n}}(S_1,w) \forall\sigma_2\in\Sol_{2^{2^n}}(S_1,\sigma_1)
\cdots
\exists\sigma_{n}\in\Sol_{2^{2^n}}(S_1,\sigma_{n-1}) : \mathsf{true}$ hold?}

\bigskip
\noindent
The idea is that the quantified solutions $\sigma_i$ represent
subcomputations of an alternating Turing-ma\-chine, where all states
in the subcomputation are either existential (if $i$ is odd) or 
universal (if $i$ is even). Our definition of vertical extension
of solutions ensures that these subcomputations can be combined into
on single computation of the alternating Turing-machine. A similar 
encoding of alternating Turing machines by tiling systems can be found in \cite{Chlebus84}.

Let $\Trans_{1}$ be the fixed GTRS graph that is obtained from 
$\Trans_{0}$ of Section \ref{S Lower} when we replace the tiling system $S_0$ by $S_1$.

\begin{cor}
The first-order theory of $\Trans_1$ is hard for $\ATIME(2^{2^{\poly(n)}},\poly(n))$
under logspace reductions.
\end{cor}

\begin{proof}
We recycle the proof presented in Section \ref{S Lower}. 
We adapt the formulas constructed in Section \ref{S Lower} to the fixed
tiling system $S_1$ (instead of $S_0$).
Recall that we can compute in logspace a formula $\sol(x)$ such that
for every  tree $t$ we have
that $\Trans_1\models\sol(t)$ if and only if $t$ corresponds
to a $2^{2^n}$-solution for $S_1$.
It is an easy exercise to construct in logspace a formula $\mathsf{ext}$ such
that for any two trees $t$ and $t'$ each satisfying $\sol$ we have
$\Trans_1\models\mathsf{ext}(t,t')$ if and only if the solution
corresponding to $t'$ extends that of $t$ vertically.
We obtain that a word $w$ (with $n=|w|$ odd) is a positive instance of the linearly
alternating $(2^{2^n}\times 2^{2^n})$ tiling problem if and only if
$\Trans_1$ is a model of the sentence
$$
\exists x_1 \left(
\sol(x)\wedge\varphi_w(x)\wedge \forall x_2
\left((\sol(x_2)\wedge \mathsf{ext}(x_1,x_2))\rightarrow\ \cdots\
\exists x_n
\left(\sol(x_n)\wedge \mathsf{ext}(x_{n-1},x_n)\right)\biggl.\right)\biggl.\right).
$$
\end{proof}
We should remark that hardness for $\ATIME(2^{2^{\poly(n)}},\poly(n))$
can be also proved using the method of Compton and Henson \cite{CH90}
(monadic interpretation of addition on large numbers).
The use of tilings has the advantage of giving an almost generic reduction.
On the other hand, the method of \cite{CH90} yields completeness
under the slightly stronger log-lin reductions.

\section{The first-order theory with regular unary predicates}
\label{sec:GTRS+unary}

For a GTRS $\R = (A,\Sigma,R)$ and a set of trees 
$L \subseteq \T_A$, we denote with $(\Trans(\R), L)$ 
the structure that results from the labelled graph $\Trans(\R)$
by adding the set $L$ as an additional unary predicate.
Note that if $L$ is a regular set of trees, then  
$(\Trans(\R), L)$ is a tree automatic structure, and hence
has a decidable first-order theory.

By the following result, our $\ATIME(2^{2^{\poly(n)}},O(n))$
upper bound for the first-order theory of a ground tree rewrite
graph does not carry over to ground tree rewrite
graphs expanded by a regular unary predicate.

\begin{thm}
There exists a fixed GTRS $\R_2=(A,\Sigma,R)$ and a fixed regular
tree language $L \subseteq \T_A$ such that the first-order theory 
of $(\Trans(\R_2), L)$ is non-elementary.
\end{thm}

\medskip

\noindent
{\em Proof sketch}.
The proof idea is an adaption of the proof of Theorem~2 in
\cite{GoLi11} and is hence only shortly sketched. 
We reduce from the satisfiability problem for first-order logic over
binary words. Binary words are considered as structures 
over the signature $(P_0,P_1,\leq)$, where $P_0$ and $P_1$ are
unary relations (representing those positions, where the
letter is $0$ and $1$, respectively), and where
$\leq$ is the natural order relation on positions.
The idea is that a tree $t \in \T_A$ (where $A_2 = \{\bullet\}$ and 
$A_0 = \{0,1\}$) corresponds to the unique word over $\{0,1\}$ that one obtains by simply 
reading the yield string
(the sequence of node labels when traversing the leaves in
lexicographic order) of $t$. 
Let $\mathsf{yield}(t)$ denote the yield string of $t$.

We translate a given first-order sentence $\varphi$ over the signature 
$(P_0,P_1,\leq)$ into a first-order formula $\psi(x)$ over the signature
of $(\Trans(\R_2), L)$ such that for every tree $t \in \T_A$ 
we have: $\mathsf{yield}(t) \models \varphi$ if and only if 
$(\Trans(\R_2), L) \models \psi(t)$.
Assume that $x_1, \ldots, x_n$ are the variables that occur in $\varphi$.
Bounding a variable $x_i$ ($1 \leq i \leq n$)
of $\varphi$ to a certain position in the word  
$\mathsf{yield}(t)$
is simulated by labelling the corresponding leaf of the tree
$t$ by a chain of unary symbols of length $i$. In order to keep the 
GTRS $\R_2$ fixed, this chain has to be built up in $i$ rewrite
steps that are controlled by the formula $\psi(x)$.
In order to verify an atomic predicate $x_i < x_j$ in the tree $t$ 
one has to check, whether the $i$-labelled node of $t$ is
lexicographically smaller than the $j$-labelled node. 
To do this using a fixed GTRS, one first replaces the 
chain of length $i$ (resp., $j$) that identifies the position to which
$x_i$ (resp., $x_j$) is bound by a special constant $a$ (resp. $b$).
Again, this process has to be controlled by the formula $\psi(x)$.
Finally, we can check $x_i < x_j$ using the regular set of trees
that contain
a unique $a$-labelled leaf and a unique $b$-labelled leaf,
and the $a$-labelled leaf is lexicographically smaller than the 
$b$-labelled leaf. This regular set will be the set $L$ in the theorem.
\qed

\section{Open problems}

We proved that the uniform first-order theory of ground tree rewrite graphs belongs to 
the complexity class $\ATIME(2^{2^{\poly(n)}},O(n))$ and that there exists a fixed ground tree
rewritie graph with an $\ATIME(2^{2^{\poly(n)}},O(n))$-complete first-order theory.

A complexity gap in this context exists for the 
first-order theory of the one-step rewrite graph of a semi-Thue system (word rewrite system):
It is known to be $2\mathsf{EXPSPACE}$-hard and decidable but it is not known to be
elementary \cite{KuLo2005}. One may try to tackle this problem using  techniques similar
to those used in this paper.

An important open problem concerning ground tree rewrite graph concerns bisimulation equivalence.
It is not known whether the following problem is decidable: Given a ground tree rewrite system $\R$ and two trees $s$ and $t$, are
$s$ and $t$ are bisimilar in the graph $\Trans(\R)$? 
For pushdown graphs this problem is decidable \cite{Senizergues05} but not elementary, as was
recently shown in \cite{BenediktGKM13}.
A further question is the complexity of deciding bisimilarity between a ground tree rewrite
system and a finite system, lying between $\mathsf{PSPACE}$ and $\mathsf{coNEXP}$ 
\cite{GoLi11}.

\section*{Acknowledgments}

The work of the second author is supported by the DFG project GELO. 
We want to thank the referees of this paper for their valuable comments.

\bibliographystyle{abbrv}

\end{document}